\documentclass[onecolumn,prd,superscriptaddress,nofootinbib,longbibliography,preprintnumbers]{revtex4-2}
\usepackage{amsmath,amsthm}
\usepackage{amssymb,amsfonts}
\usepackage{bm}
\usepackage[colorlinks=true, linkcolor=blue, citecolor=blue, urlcolor=blue]{hyperref}
\usepackage{graphicx}% Include figure files
\usepackage{braket}
\usepackage{subfigure}
\usepackage{cases}
\usepackage{multirow}
\numberwithin{equation}{section}

\usepackage{algpseudocode}
\usepackage{fancyhdr}
\usepackage{setspace}

\newcounter{algorithm}
\renewcommand{\thealgorithm}{\arabic{algorithm}}

\newenvironment{myalgorithm}[1][]{
  \refstepcounter{algorithm}%
  \par\medskip
  \noindent\textbf{Algorithm \thealgorithm. #1}\par
  \vspace{0.5\baselineskip}
}{%
  \par\medskip
}

\newcommand{\im}{\mathrm{i}}
\newcommand{\diff}{\mathrm{d}}

\newcommand{\Tr}{\mathrm{Tr}}

\newcommand{\e}{\mathrm{e}}

\newcommand{\rme}{\mathrm{e}}

\usepackage{tikz,xcolor}
\definecolor{lime}{HTML}{A6CE39}
\DeclareRobustCommand{\orcidicon}{%
	\begin{tikzpicture}
	\draw[lime, fill=lime] (0,0) 
	circle [radius=0.16] 
	node[white] {{\fontfamily{qag}\selectfont \tiny ID}};	\draw[white, fill=white] (-0.0625,0.095) 
	circle [radius=0.007];	\end{tikzpicture}
	\hspace{-2mm}}
\foreach \x in {A,B,C,D,E}{%
	\expandafter\xdef\csname orcid\x\endcsname{\noexpand\href{https://orcid.org/\csname orcidauthor\x\endcsname}{\noexpand\orcidicon}}
}

\newcommand{\FNAL}{\affiliation{
Fermi National Accelerator Laboratory, Batavia, Illinois, USA}}

\newcommand{\Cone}{\affiliation{Capital One, Chicago, Illinois, USA}}

\newcommand{\NCSU}{\affiliation{
Department of Physics and Astronomy, North Carolina State University, Raleigh, North Carolina 27695, USA}}

\newcommand{\TU}{\affiliation{Center for Computational Sciences, University of Tsukuba, Tsukuba, Ibaraki 305-8577, Japan}}

\newcommand{\UTokyo}{\affiliation{Graduate School of Science, The University of Tokyo, Bunkyo-ku, Tokyo, 113-0033, Japan}}

\newcommand{\KU}{\affiliation{Department of Physics,
Kyoto University, Kyoto, 606-8502, Japan}}

\begin{document}

\title{Tensor renormalization group approach to critical phenomena via symmetry-twisted partition functions}

\author{Shinichiro Akiyama~\orcidA{}}
\email{akiyama@ccs.tsukuba.ac.jp}
\TU
\UTokyo

\author{Raghav G. Jha~\orcidB{}}
\email{raghav.govind.jha@gmail.com}
\NCSU

\author{Jun Maeda~\orcidC{}}
\email{maeda@gauge.scphys.kyoto-u.ac.jp}
\KU

\author{Yuya Tanizaki~\orcidD{}}
\email{yuya.tanizaki@yukawa.kyoto-u.ac.jp}
\affiliation{Yukawa Institute for Theoretical Physics, Kyoto University, Kyoto 606-8502, Japan}

\author{Judah Unmuth-Yockey~\orcidE{}}
\email{jfunmuthyockey@gmail.com}
\Cone\FNAL

\preprint{
    UTHEP-814, UTCCS-P-172, KUNS-3083, 
    YITP-25-180
}

\begin{abstract}
    The locality of field theories strongly constrains the possible behaviors of symmetry-twisted partition functions, and thus they serve as order parameters to detect low-energy realizations of global symmetries, such as spontaneous symmetry breaking (SSB). 
    We demonstrate that the tensor renormalization group (TRG) offers an efficient framework to compute the symmetry-twisted partition functions, which enables us to detect the symmetry-breaking transition and also to study associated critical phenomena. 
    As concrete examples of SSB, we investigate the two-dimensional (2D) classical Ising model and the three-dimensional (3D) classical $O(2)$ nonlinear sigma model, and we identify their critical points solely from the twisted partition function. By employing the finite-size scaling argument, we find the critical temperature $T_c=2.2017(2)$ with the critical exponent $\nu = 0.663(33)$ for the 3D $O(2)$ model. 
    In addition, we also study the Berezinskii--Kosterlitz--Thouless (BKT) criticality of the 2D classical $O(2)$ model by extracting the helicity modulus from the twisted partition functions, and we obtain the BKT transition temperature, $T_{\mathrm{BKT}}=0.8928(2)$. 
\end{abstract}

\maketitle

\onecolumngrid

\section{Introduction}

Tensor networks have become a powerful framework for investigating quantum many-body physics. One of the most successful approaches is the density matrix renormalization group (DMRG)~\cite{White:1992zz}, which is a variational method based on the matrix product state~\cite{Schollw_ck_2011}. 
In parallel, tensor networks also provide a means to study quantum systems through their corresponding classical counterparts in the path-integral formalism. The corner transfer matrix renormalization group (CTMRG)~\cite{Nishino_1996} is one such method. 
It was pointed out in Ref.~\cite{Nishino_1997} that CTMRG can be understood as a variational approach grounded in Baxter’s corner transfer matrix formulation~\cite{Baxter:1968krk,baxter1978variational}.
Furthermore, the tensor renormalization group (TRG)~\cite{Levin:2006jai} constitutes a practical realization of real-space renormalization group methods, allowing for numerical evaluation of partition functions and their path-integral representations.

By directly computing partition functions and path integrals, the TRG gives access to a wide range of thermodynamic quantities, which are crucial for understanding phase structures of many-body systems.
To this end, several techniques, such as the impurity tensor method~\cite{MORITA201965} and automatic differentiation~\cite{PhysRevX.9.031041}, have been developed to compute thermodynamic quantities with high accuracy.
In addition, the TRG also offers an approach to studying critical phenomena via conformal field theory (CFT) data, such as central charges and scaling dimensions, owing to its capability to numerically construct transfer matrices~\cite{PhysRevB.80.155131}, and various TRG algorithms including the tensor network renormalization (TNR) algorithm~\cite{PhysRevLett.115.180405,PhysRevLett.118.110504,Iino:2019rxt,Homma_2024,ueda2025globaltensornetworkrenormalization} have been widely used to accurately access the CFT data.

Gu and Wen proposed a way to determine critical points through a ratio of partition functions with periodic boundary conditions~\cite{PhysRevB.80.155131}, which we refer to as the Gu--Wen ratio in this paper.
This ratio computes the ground-state degeneracy, just from partition functions, when the system has a discrete global symmetry. 
Therefore, the Gu--Wen ratio has been used to locate the critical point associated with the spontaneous $\mathbb{Z}_{N}$ symmetry breaking involving various spin models~\cite{Wang_2014,Chen:2017ums,Akiyama:2019xzy,PhysRevResearch.4.023159}, the $\mathbb{Z}_{2}$ charge conjugation symmetry breaking in the lattice gauge theories with a topological $\theta$ term~\cite{Akiyama:2024qer,Kanno:2024elz}, and confinement-deconfinement phase transition at finite temperature in the $\mathbb{Z}_{2}$ gauge theory~\cite{Kuramashi:2018mmi}.
In addition to the finite-size scaling analysis for the Gu--Wen ratio~\cite{Morita:2024lwg}, the universal values for the Gu--Wen ratios at criticality can also be obtained from the CFT through the modular-invariant partition functions~\cite{Aizawa:2025lxi,Morita:2025hsv}.
Furthermore, recent work~\cite{Akiyama:2024qgv} has suggested that the Gu--Wen ratio may also be useful for locating critical points associated with the spontaneous breaking of global continuous symmetries. 

The essence of the Gu--Wen ratio is based on the fact that the locality of field theories strongly constrains the possible behaviors of the low-energy effective theory and its partition function. 
The partition function itself contains the non-universal contribution that depends on the details of the ultraviolet (UV) physics, but this ambiguity appears only through the local counter-term of the low-energy effective theory. 
The Gu--Wen ratio is designed to eliminate this non-universal contribution and extract the low-energy data, such as the ground-state degeneracy for the spontaneously broken discrete symmetries. 

We can now readily extend this idea to the symmetry-twisted partition functions when the field theory of interest enjoys global discrete and/or continuous symmetries~\cite{Maeda:2025ycr} (see Refs.~\cite{Bedaque:2004kc, Sachrajda:2004mi} for earlier works of using symmetry twists in lattice QCD). 
In this paper, we demonstrate that the ratios between the symmetry-twisted and ordinary partition functions are useful order parameters to identify the phase transitions associated with the spontaneous symmetry breaking (SSB) for both discrete and continuous global symmetries. 
We show that the TRG approach is suited to compute the twisted partition functions by imposing the symmetric structure in tensor networks. 
For our demonstration, we numerically compute the twisted partition functions of the two-dimensional (2D) Ising model and also of the three-dimensional (3D) $O(2)$ model, and their critical phenomena are successfully studied. 
We further apply our method to the 2D $O(2)$ model, where our approach enables a direct computation of the helicity modulus from the twisted partition functions, and we get a clear signal for the Berezinskii--Kosterlitz--Thouless (BKT) transition.

The paper is organized as follows. In Sec.~\ref{sec:TwistedPartitionFunction}, we give a brief review of the symmetry-twisted partition functions and their universal behavior based on Ref.~\cite{Maeda:2025ycr}. 
We then explain the way to compute the twisted partition functions using the TRG method in Sec.~\ref{sec:TRGmethod_TwistedZ}. 
In Sec.~\ref{sec:NumericalResults}, we show our numerical results of the twisted partition functions for the 2D Ising model (for $\mathbb{Z}_2$ SSB), the 3D $O(2)$ model (for $U(1)$ SSB), and also the 2D $O(2)$ model (for BKT criticality). 
Conclusions and discussions are given in Sec.~\ref{sec:Conclusion}. 
In Appendix~\ref{appendix:CFT_twistedZ}, we give the general computation of the twisted partition functions on the torus for 2D CFTs. In Appendix~\ref{appendix:3DXY}, we discuss the finite bond-dimension effect of the 3D $O(2)$ model in the thermodynamic limit.

\section{Twisted Partition Functions as Order Parameters for Spontaneous-Symmetry Breaking}
\label{sec:TwistedPartitionFunction}

Let us consider a local field theory that describes a Euclidean quantum field theory (QFT) or a classical statistical system. We assume that the field theory has a global symmetry, and we would like to diagnose its low-energy realization. 
The most standard approach to this problem is to consider correlation functions of the local order parameters. Their off-diagonal long-range behavior detects the spontaneous breaking of the global symmetry, and this technique is quite powerful when we study the system numerically using the Monte Carlo method. 
For the TRG, however, it is not necessarily easy to compute the long-range behaviors of local correlators, and some alternative quantities would be desired to efficiently identify the low-energy symmetry realization using TRG.

With global symmetries of local field theories, one can introduce their background gauge fields, which we formally write as $A$, and let $Z[A]$ denote the partition function under the presence of the background gauge field $A$. 
To be concrete, let us take a lattice regularization of the system of interest, and we further assume that the global symmetry $G$ without 't~Hooft anomalies is realized as the on-site action. For the lattice degrees of freedom $\phi_x$, where $x$ labels a site on the lattice $\Gamma$, the symmetry $g\in G$ acts as $\phi_x\mapsto g\cdot \phi_x$. 
Then, the background gauge field $A$ is nothing but the set of link variables $\{U_{x,\mu}\}_{x,\mu}\in G^{\otimes(\mathrm{links})}$, where we replace the hopping term $\phi^\dagger_{x+\hat{\mu}}\phi_x$ by $\phi^\dagger_{x+\hat{\mu}}U_{x,\mu}\phi_x$. 
This achieves the local gauge invariance for the background gauge field, 
\begin{equation}
    Z[A^\lambda]=Z[A], 
    \label{eq:bkg_gauge_invariance}
\end{equation}
where $\lambda:\Gamma\to G$ is the gauge-transformation parameter and $A^\lambda=\{\lambda_{x+\mu}^{-1}U_{x,\mu}\lambda_{x}\}_{x,\mu}$ is the gauge transformation of $A$ by $\lambda$. 
This relation is exact by construction, and thus it has to be respected by the low-energy effective field theory. 
This fact gives a strong constraint on the low-energy effective theory and its coupling to the background field $A$. 

In this paper, we focus on the cases where the background gauge field $A$ is flat, i.e., $\mathcal{P}\prod_{(x,\mu)\in \partial p} U_{x,\mu}=1 \in G$ for any local plaquettes $p$. 
Then, the only gauge-invariant data of the background field $A$ is captured by its holonomies around nontrivial cycles $\gamma$, i.e., $\mathcal{P}\exp(i \int_{\gamma} A)$, and these holonomies correspond to turning on symmetry-twisted boundary conditions, which can be explicitly seen by eliminating $A$ from the Lagrangian via an aperiodic gauge transformation. 
The exact property~\eqref{eq:bkg_gauge_invariance} combined with the locality strongly restricts possible behaviors of the symmetry-twisted partition functions, i.e. $Z[A]$ for flat $A$ fields, and this justifies to use $Z[A]$ as the order parameters. In the following part of this section, we shall review its properties based on Ref.~\cite{Maeda:2025ycr} for some basic examples that are relevant in the rest of this paper.\footnote{We note that the idea of using the symmetry-twisted partition function itself is not a new one, and see, e.g., Refs.~\cite{HASENBUSCH1993423, Hukushima_1999, Campostrini:2000iw, deForcrand:2000fi} for Monte Carlo studies on computing the symmetry-twisted partition functions.} 
In later sections, we will show that these quantities can be efficiently computed using the TRG techniques.

%%%%%%%%%%%%%%%%%%%

\subsection{Spontaneous breaking of discrete symmetries}

Let us consider a gapped phase where a discrete symmetry $G$ is spontaneously broken to a subgroup $H$, $G\xrightarrow{\mathrm{SSB}}H$. 
We take the Euclidean spacetime as the torus $T^d$ of the size $L_1\times L_2\times \cdots \times L_d$. 
The flat background gauge field $A$ is characterized by the holonomies $(g_1,g_2,\ldots, g_d)$ along each direction of $T^d$, and we completely identify them in the following. 
Here, let us take the symmetry twist only along the $d$-th direction, $A=(1,\ldots, 1, g_0)$, and we consider the infinite-volume behavior of the partition function $Z_{g_0}=Z[A=(1,\ldots, 1, g_0)]$ with the symmetry twist by $g_0\in G$.  

Since we have assumed the presence of the mass gap, the low-energy limit of the system is characterized by the label of the vacua satisfying the cluster-decomposition property. 
For $G\xrightarrow{\mathrm{SSB}}H$, each vacuum state is labeled by an element of the left coset $G/H=\{gH \mid g\in G\}$~\cite{Coleman:1969sm, Callan:1969sn}. 
When we impose a twisted boundary condition with $g_0 \in G$ along the $d$-th direction, a vacuum $gH$ is mapped to $g_0gH$ under the twist. Thus, the dominant contribution to the twisted partition function comes from the vacuum states that remain invariant under the twist, i.e., those satisfying $g_0gH=gH$. If no such invariant vacuum exists, the configuration necessarily contains a domain wall separating different vacua. Since a domain wall carries a finite tension $\Lambda_{\mathrm{DW}}$ and extends over a spatial volume, its contribution to the partition function gives exponential suppression as $\exp(-\Lambda_{\mathrm{DW}} L_1\cdots L_{d-1})$ in the large-volume limit and becomes negligible. 
Therefore, the twisted partition function $Z_{g_0}$ in the large-volume limit behaves as
\begin{align}
    Z_{g_0} = |\{gH \in G/H \mid g^{-1}g_0g\in H\}| \times \rme^{-\int_{T^d} \Lambda \,\diff^d x}, 
\end{align}
where the prefactor $|\{gH \in G/H \mid g^{-1}g_0g\in H\}|$ gives the universal low-energy data that counts the number of invariant vacua by the $g_0$ action (i.e., the number of vacua on the symmetry-twisted torus without domain walls), 
while $\Lambda$ describes the ground-state energy density (or the cosmological constant), which is non-universal and depends on the details of the UV regularization. 
To extract the universal data of the symmetry breaking, we have to take suitable ratios of these partition functions that become independent of $\Lambda$. 
In Ref.~\cite{PhysRevB.80.155131}, Gu and Wen proposed the following quantity, 
\begin{equation}
    \frac{Z_{1}(L_1\times \cdots \times L_{d-1}\times L_d)^2}{Z_{1}(L_1\times \cdots \times L_{d-1}\times 2L_d)} = |G/H|, 
\end{equation}
which is often called the Gu--Wen ratio and detects the number of degenerate vacua for $G\xrightarrow{\mathrm{SSB}}H$ using untwisted partition functions with different spacetime volumes. 

Here, we instead consider the ratio of the partition functions with the same spacetime volume but with different symmetry twists,
\begin{equation}
    \frac{Z_{g_0}(L_1\times \cdots \times L_d)}{Z_{1}(L_1\times \cdots \times L_d )} 
    = \frac{|\{gH \in G/H \mid g^{-1}g_0g\in H\}|}{|G/H|}. 
\end{equation}
Especially when $H$ is a normal subgroup (i.e., $g^{-1}H g = H$ for any $g\in G$), this expression simplifies to
\begin{align} \label{eq:behavior_of_Zg}
    \frac{Z_{g_0}}{Z_{1}} = 
    \begin{cases}
        1 & g_0 \in H \\
        0 & g_0 \notin H. \\
    \end{cases}
\end{align}
Thus, the ratio of the symmetry-twisted partition function tells the details of the unbroken subgroup $H$ itself, while the Gu--Wen ratio computes its size, $|H|$. 
If the symmetry is not spontaneously broken at all, we should have $Z_{g_0}/Z_{1}=1$ for any $g_0\in G$, and thus the sudden drop of the twisted partition function identifies the location of the symmetry-breaking phase transition.

We now find that the ratio $Z_{g_0}/Z_{1}$ for $g_0\not\in H$ jumps from $1$ to $0$ across the phase transition from the symmetric to symmetry-broken phases (Here, we assume that $H$ is a normal subgroup of $G$ just for simplicity). 
If this phase transition is second order, we can obtain the universal curve for $Z_{g_0}/Z_1$ using the finite-size scaling argument. 
For simplicity, let us first consider the case of the $d$-dimensional torus with the length $L\,(=L_1=\cdots =L_d)$. 
We assume that the critical point is described by a conformal field theory (CFT)  $S_{\mathrm{CFT}}$, 
and we consider its relevant deformation that corresponds to the temperature perturbation,
\begin{align}
    S = S_{\rm CFT} + t\int \diff^dx\,\Phi,
\end{align}
where the scaling dimension of $\Phi$ is given by $\Delta$. In this case, the mass scale $m$ induced by the perturbation behaves as $m\sim t^{1/(d-\Delta)}$, so that the only dimensionless combination involving the system size is $tL^{1/\nu}$ with $\nu=1/(d-\Delta)$. Since $Z_{g_0}/Z_1$ is a dimensionless quantity, its finite-size scaling form must be given by a universal function of this parameter:
\begin{align}
    \frac{Z_{g_0}(L\times \cdots\times L)}{Z_1(L\times \cdots \times L)} = f(tL^{1/\nu}).
\end{align}
For rectangle boxes, the universal form becomes 
\begin{align}
    \frac{Z_{g_0}(L_1\times \cdots\times L_d)}{Z_1(L_1\times \cdots \times L_d)} = \tilde{f}(tL_d^{1/\nu}, L_1/L_d,\cdots, L_{d-1}/L_d).
\end{align}
At the critical point (i.e., $t=0$), the value of this ratio is completely determined by the underlying CFT.

%%%%%%%%%%%%%%%%%%%

\subsection{Spontaneous breaking of \texorpdfstring{$U(1)$}{U(1)} symmetry}

Next, let us discuss the behavior of the twisted partition function in the SSB phase of $U(1)$ symmetry. 
We consider the torus $T^d$ of size $L_1\times \cdots \times L_d$ with the flat background $U(1)$ gauge field with the holonomies $(1,\ldots, 1,\rme^{\im \alpha})$. 
We assume $d\ge 3$ since the Mermin--Wagner--Coleman theorem~\cite{Mermin:1966fe,Coleman:1973ci} forbids the spontaneous breaking of continuous symmetry in two dimensions or lower.

To be specific, we consider a complex scalar field theory,
\begin{align}
    S[\phi, A] = \int \diff^dx \left[ |(\partial_{\mu}+\im A_\mu)\phi|^2+m^2|\phi|^2 + \frac{\lambda}{4}|\phi|^4 \right].
\end{align}
This theory has a $U(1)$ symmetry of phase rotation, and its background gauge field is taken as $A=\alpha \frac{\diff x_d}{L_d}$. We can perform the aperiodic gauge transformation that eliminates $A$ from the Lagrangian, but it imposes the twisted boundary condition, 
\begin{equation}
    \phi(x_1,\ldots, x_{d-1}, x_d+L_d)=\rme^{\im \alpha} \phi(x_1,\ldots, x_{d-1},x_d). 
\end{equation}
Other directions are periodic without any twists. 
Let us write the $U(1)$-twisted partition function as $Z_\alpha:=Z[A]$ instead of $Z_{\rme^{\im \alpha}}$ for better readability. 

In the trivial phase (i.e. $m^2>0$), the potential minimum $\phi(x)=0$ is consistent with the twist and therefore the ratio of the partition function behaves as 
\begin{align}\label{eq:zpz_trivial_phase}
    \frac{Z_{\alpha}}{Z_{0}} = 1, 
\end{align}
when $L_d$ is sufficiently large compared to the inverse mass gap.
% RGJ: Fixed sentence here

In the symmetry-broken phase (i.e., $m^2=-\frac{1}{4}\lambda v^2<0$), non-zero twist changes a vacuum state to another vacuum state. 
By parametrizing the classical minima as $\phi(x)=\frac{v}{\sqrt{2}}\, \rme^{\im \theta(x)}$, the low-energy effective Lagrangian becomes 
\begin{equation}
    S_{\mathrm{eff}}[\theta,A]=\int \diff^d x \frac{v^2}{2}(\partial_\mu \theta+A_\mu)^2
    \label{eq:effectiveLagrangian_U1SSB}
\end{equation} 
at the leading order. If we eliminate $A_\mu$ by the aperiodic gauge transformation, the twisted boundary condition becomes $\theta(x_1,\ldots, x_{d-1}, x_d+L_d)=\theta(x_1,\ldots, x_{d-1}, x_d)+\alpha$ mod $2\pi$. 
%Therefore, the field configuration must interpolate two states\footnote{Note that this field configuration is not sharply localized domain wall as opposed to the case of discrete symmetry.}. 
The twisted partition function with this effective Lagrangian can be calculated as
\begin{align}
    \frac{Z_{\alpha}(L_1\times \ldots\times L_{d-1}\times L_d)}{Z_{0}(L_1\times \ldots\times L_{d-1}\times L_d)} 
    = \frac{\sum_w \exp\left[-\frac{v^2 L_1\cdots L_{d-1}}{2L_d}(2\pi w-\alpha)^2\right]}{\sum_{w'} \exp\left[-\frac{v^2 L_1\cdots L_{d-1}}{2L_d}(2\pi w')^2\right]}.
    \label{eq:twistedZ_U1SSB}
\end{align}
In the large-volume limit, the contribution from $w=0$ dominates for $-\pi < \alpha < \pi$, while the contribution from $w=1$ dominates for $\pi<\alpha<3\pi$. 
At $\alpha=\pi$, contributions from $w=0, 1$ become equal. 
In all the cases with nontrivial twists $\alpha\not\in 2\pi \mathbb{Z}$, the ratio of the partition function approaches zero in the large-volume limit, $L_{i=1,\ldots,d}\to \infty$ with fixed $L_i/L_j$, for $d\ge 3$. 

Therefore, as in the case of a discrete symmetry, the ratio of the partition function also can be used to diagnose the trivial phase and the SSB phase: It converges to $1$ for the symmetric phase and to $0$ for the SSB phase in the large-volume limit, so the criterion for the symmetry breaking is crystal-clear.\footnote{This is an important difference between the symmetry-twisted partition function and the Gu--Wen ratio for the cases of continuous symmetries. 
In the $U(1)$ SSB phase, the first natural guess would tell the Gu--Wen ratio should converge to $2\pi$, which comes from the integration over the classical moduli. 
For the case of the discrete symmetry breaking, such an expectation based on the classical analysis gives the correct answer because the system is gapped. 
With the presence of massless Nambu--Goldstone modes, however, the Gu--Wen ratio gets further corrections as their fluctuation determinants in the numerator and denominator of the Gu--Wen ratio do not cancel, and it requires some computations to judge if the system is in the broken phase from the value of the Gu--Wen ratio. } 
When the symmetry-breaking transition is of the second order, the finite-size scaling of $Z_{\alpha}/Z_0$ near the critical point can also be discussed in a similar manner as in the case of a discrete symmetry.

We shall make a brief comment on the 2D case. 
In this case, the superfluid vortices behave as instantons, which can potentially restore the classically broken $U(1)$ symmetry. 
To take their effects into account, it is convenient to take the Abelian duality, $v^2\partial_\mu \theta \Leftrightarrow \frac{\im}{2\pi}\varepsilon_{\mu\nu}\partial_\nu \tilde{\theta}$, so that the vortex insertion can be easily described by a local operator $\e^{\im \tilde{\theta}}$. 
Then, the dilute gas approximation of vortices gives the effective action for the dual compact boson as 
\begin{align}
    S_{\mathrm{eff}}[\tilde{\theta},A]=\int \diff^2x\left[\frac{R^2}{4\pi}(\partial_{\mu}\tilde{\theta})^2 -\mu\cos\tilde{\theta}\right]+\frac{\im}{2\pi}\int A \wedge \diff \tilde{\theta},
\end{align}
where $2\pi v^2 R^2=1$ and $\mu$ is the vortex fugacity. When the cosine term is relevant, the system exhibits the trivially gapped phase, and the ratio of the partition function behaves as~\eqref{eq:zpz_trivial_phase}. On the other hand, when the cosine term is irrelevant, the system flows to the compact boson CFT, where the value of $Z_{\alpha}/Z$ remains finite in the large-volume limit:\footnote{In \eqref{eq:twistedZ_U1SSB}, $w\in \mathbb{Z}$ denotes the winding number of $\theta$ along the twisted (or $d$-th) direction, but in \eqref{eq:twistedZ_BKTsuperfluid}, it appears as the dual variable via the Poisson resummation for the winding number of the dual boson $\tilde{\theta}$ along the periodic (i.e., untwisted) direction.}
\begin{equation}
    \frac{Z_{\alpha}(L_1\times L_2)}{Z_{0}(L_1\times L_2)}= 
    \frac{\sum_w \exp\left[-\frac{L_1}{L_2}\frac{1}{4\pi R_{\mathrm{ren}}^2}(2\pi w-\alpha)^2\right]}{\sum_{w'} \exp\left[-\frac{L_1}{L_2}\frac{1}{4\pi R_{\mathrm{ren}}^2}(2\pi w')^2\right]},
    \label{eq:twistedZ_BKTsuperfluid}
\end{equation}
where $R_{\mathrm{ren}}$ is the renormalized value of the dual compact-boson radius. 
We should note that this ratio survives in the infinite-volume limit with fixed $L_1/L_2$, unlike the case of $d\ge 3$, which is consistent with the fact that the $U(1)$ symmetry cannot be spontaneously broken due to the Mermin--Wagner--Coleman theorem~\cite{Mermin:1966fe, Coleman:1973ci}. 
Instead of $U(1)$ SSB, the low-temperature regime is described by the compact boson CFTs that explain the algebraic long-range order. 
The high-temperature gapped phase and the low-temperature conformal phase are separated by the Berezinskii--Kosterlitz--Thouless (BKT) transition~\cite{Berezinskii:1972fet, Kosterlitz:1973xp}, which happens when the vortex term crosses the marginality bound, $R_{\mathrm{ren}}^2=1/4$. 
As the BKT transition is not caused by the relevant perturbation, unlike previous examples, the finite-size scaling there gets complicated logarithmic corrections (see, e.g., Ref.~\cite{Pelissetto:2012gv} and references therein for details). 

%%%%%%%%%%%%%%%%%%%

\section{TRG approach for the symmetry-twisted partition function}
\label{sec:TRGmethod_TwistedZ}

In the previous section, we have seen that the symmetry-twisted partition function plays the role of the order parameter. 
In the Monte Carlo simulation, however, the standard approach of using the local order parameter is often more efficient than using the twisted partition functions. 
To evaluate $Z_{g_0}/Z_{1}$ in the Monte Carlo method, we should compute the expectation value of the extended observable $\exp(-(S[\phi,A]-S[\phi]))$, 
\begin{eqnarray}
    \frac{Z[A]}{Z[0]}&=&\frac{\int \mathcal{D}\phi\, \rme^{-S[\phi]}\exp(-(S[\phi,A]-S[\phi])) }{\int \mathcal{D}\phi\, \rme^{-S[\phi]}} \notag\\
%    \Bigl\langle \rme^{-(S[\phi,A]-S[\phi])} \Bigr\rangle_{\rme^{-S[\phi]}}
    &\approx& \frac{1}{N_{\mathrm{sample}}}\sum_{s=1}^{N_{\mathrm{sample}}} \exp(-(S[\phi_s,A]-S[\phi_s])),
\end{eqnarray}
with the ensembles $\{\phi_s\}_{s=1,\ldots,N_{\mathrm{sample}}}$ generated by the Boltzmann weight $\exp(-S[\phi])$. 
Since the location with nontrivial $A$ spreads over a codimension-$1$ surface that wraps around the spacetime, one can easily imagine that this quantity suffers from the severe overlap problem. 
One way to evade this overlap problem is to apply the reweighting method repeatedly so that the defect locations of the nontrivial $A$ field are gradually enlarged, and then one can obtain $Z_{g_0}/Z_{1}$ by multiplying those reweighting factors~\cite{deForcrand:2000fi}. 
Since the evaluation of the reweighting factor at each step would have roughly the same amount of computational cost of evaluating local operators, the direct measurement of the local order parameters is practically more efficient than computing the twisted partition functions in the Monte Carlo method to detect SSB.\footnote{We should note, however, that local order parameters do not fully capture quantum gapped phases. 
To detect other quantum phases, such as symmetry-protected topological (SPT) states, local operators do not have characteristic behaviors and the evaluation of the disorder operator (i.e., $Z[A]$ with non-flat $A$) and/or the twisted partition functions (i.e., $Z[A]$ with flat $A$) become essential~\cite{tHooft:1977nqb, tHooft:1979rtg, kennedy1992hidden, Kennedy:1992ifl, oshikawa1992hidden, Li:2023ani, Nguyen:2023fun}. }

With the TRG method, on the other hand, the computation of the twisted partition function turns out to be quite straightforward, at least on the formulation level. 
Let us begin with the evaluation of the usual partition function. 
When we employ the TRG algorithms to compute partition functions, we first need to express the partition functions as tensor networks and their contraction.
When the original theory is translationally invariant on a lattice, a uniform tensor network representation is obtained, i.e., the tensor $T_x$ at each site $x$ on the lattice $\Gamma$ takes the identical functional form. 
Let us formally write the tensor contraction operation as
\begin{align}
\label{eq:part_TN}
    Z
    =
    {\rm tTr}
    \left[
        \prod_{x\in\Gamma}T_{x}
    \right]. 
\end{align}
The TRG algorithms then approximately carry out these tensor contractions, which compress the information of the tensor network $\prod_x T_x$ efficiently and then take its tensor trace, $\mathrm{tTr}$.

When the theory possesses an on-site global symmetry $G$, it is possible to construct a local tensor $T_{x}$ such that $T_{x}$ encodes the constraint imposed by the original symmetry~\cite{Liu:2013nsa,Akiyama:2020sfo}: The basic idea is to decompose the degrees of freedom by the irreducible representations of $G$, and then the bond degrees of freedom also enjoy the decomposition in terms of the representations and their fusion rule is described by the Clebsch--Gordan coefficients.
This constraint can be interpreted as a discrete analogue of the Noether current conservation law~\cite{Meurice:2019ddf}.
In the case of lattice gauge theories, it can be viewed as the Maxwell equations in the tensor formulation~\cite{Meurice:2020gcd}.
A detailed discussion on the tensor network formulation can be found in Refs.~\cite{Meurice:2020pxc,Akiyama:2024ush}.
Then, in the last step of the tensor trace to obtain the partition function, we just need to multiply the character for each representation for the twisted partition function, which is denoted by $\mathrm{tTr}_{g_0}$:
\begin{equation}
    Z_{g_0}=\mathrm{tTr}_{g_0}
    \left[
        \prod_{x\in\Gamma}T_{x}
    \right]. 
\end{equation}
The intermediate steps of the TRG method, which compress the tensor $\prod_x T_x$, do not need to be changed at all as long as our contraction algorithm keeps track of the information of the on-site global symmetry.

In numerical computations of this paper, we focus on the examples with Abelian global symmetries. 
In those cases, the fusion channel of representations is always unique, which simplifies the algorithm in practice: 
We then implement the TRG with symmetry blocking~\cite{Yang:2015rra}, where the conservation law is explicitly enforced.
This symmetry blocking allows us to impose twisted boundary conditions with a finite twist angle in a sector-by-sector manner.
We sketch the procedure to compute this in Algorithm~\ref{alg:TRGalgo1}.

\begin{myalgorithm}[Computation of $Z_{-1}/Z_{1}$ (corresponding to $\pi$-twist) for the 2D classical Ising model]
    \begin{algorithmic}[1]
    \setstretch{1.3} 
        \State Construct the rank-4 local tensor $T_{ijkl}$. The construction ensures the $\mathbb{Z}_{2}$ symmetry: only elements with $(i+j+k+l) \bmod 2 = 0$ are non-zero. Create a matrix $M^{(n)}$ where the $\mathbb{Z}_{2}$-even is the upper left block and $\mathbb{Z}_{2}$-odd is the lower right block and $n$ denotes the coarse-graining step. Reshape the matrix back to the rank-4 tensor. 
        \State Perform the singular value decomposition on each block separately and carry out a coarse-graining step with a given TRG algorithm. Reshape the tensor back to a matrix.
        \State  Compute the normal trace of the entire matrix, $M^{(n)}$ which gives $Z_{1} = \operatorname{Tr} M^{(n)}_{\mathrm{even}}
        + \operatorname{Tr} M^{(n)}_{\mathrm{odd}}$. 
        $Z_{-1}$ is computed by taking the normal trace for the $\mathbb{Z}_{2}$-even block and that for the $\mathbb{Z}_{2}$-odd block, with adding a factor of $-1$ i.e., $Z_{-1} = \operatorname{Tr} M^{(n)}_{\mathrm{even}}
        - \operatorname{Tr} M^{(n)}_{\mathrm{odd}}$. Then compute the ratio $Z_{-1}/Z_{1}$. 
        \State Repeat steps 1--3 to obtain 
        $Z_{-1}/Z_{1}$ on a lattice of size $L\times L$, where $L = 2^n$. 
    \end{algorithmic}
    \label{alg:TRGalgo1}
\end{myalgorithm}
We note that the computational cost is the same as the given TRG algorithm.

%%%%%%%%%%%%%%%%%%%

\section{Numerical Results}
\label{sec:NumericalResults}

In this section, we show the numerical results of the symmetry-twisted partition functions across the symmetry-breaking phase transitions for several classical statistical models. For the 2D models, we use the bond-weighted tensor renormalization group (BTRG) algorithm~\cite{PhysRevB.105.L060402} while for the 3D classical spin model, we use the anisotropic tensor renormalization group (ATRG) algorithm~\cite{Adachi:2019paf}.

First, to demonstrate that twisted partition functions work correctly as order parameters, we apply our approach to the 2D classical Ising model as a benchmark, and our method and the conventional Gu--Wen ratio equally work well in this model.
We next consider the 3D classical $O(2)$ model as a nontrivial example, where the Gu--Wen ratio cannot be used to locate the critical point. 
We shall see that the symmetry-twisted partition function gives a clear signal of the phase transition. 

Lastly, we consider the 2D classical $O(2)$ model, where the BKT transition happens. Using the twisted partition function, we extract the helicity modulus (superfluid stiffness) using the BTRG algorithm. The helicity modulus, introduced by Fisher, Barber, and Jasnow, effectively measures the response of a system to a `twist'~\cite{Fisher1973}, and thus our computation closely follows the original definition of the helicity modulus. 

\subsection{2D Ising model}
\label{sec:NumericalResults_2DIsing}

Let us consider the classical Ising model, 
\begin{equation}
    S[\sigma_x]=-\beta \sum_{\langle xy\rangle}\sigma_x \sigma_y, 
\end{equation}
where $\sigma_x\in \{\pm 1\}$ denotes the $\mathbb{Z}_2$ spin and $T=1/\beta$ is the temperature. 
There exists on-site $\mathbb{Z}_2$ symmetry i.e., $\sigma_x\mapsto -\sigma_x$, and we discuss its spontaneous breaking: At low temperatures $T<T_c$, the twisted partition function should go to $0$, $Z_{-1}(L\times L)/Z_{1}(L\times L)\to 0$, as $L\to \infty$, while at high temperatures $T>T_c$, $Z_{-1}(L\times L)/Z_{1}(L\times L)\to 1$. 
This model is exactly solvable, and the location of the symmetry-breaking phase transition is at $T_{c}=2/(\log(1+\sqrt{2}))$ for a square lattice. 

\begin{figure}
    \centering
    \includegraphics[width=12cm]{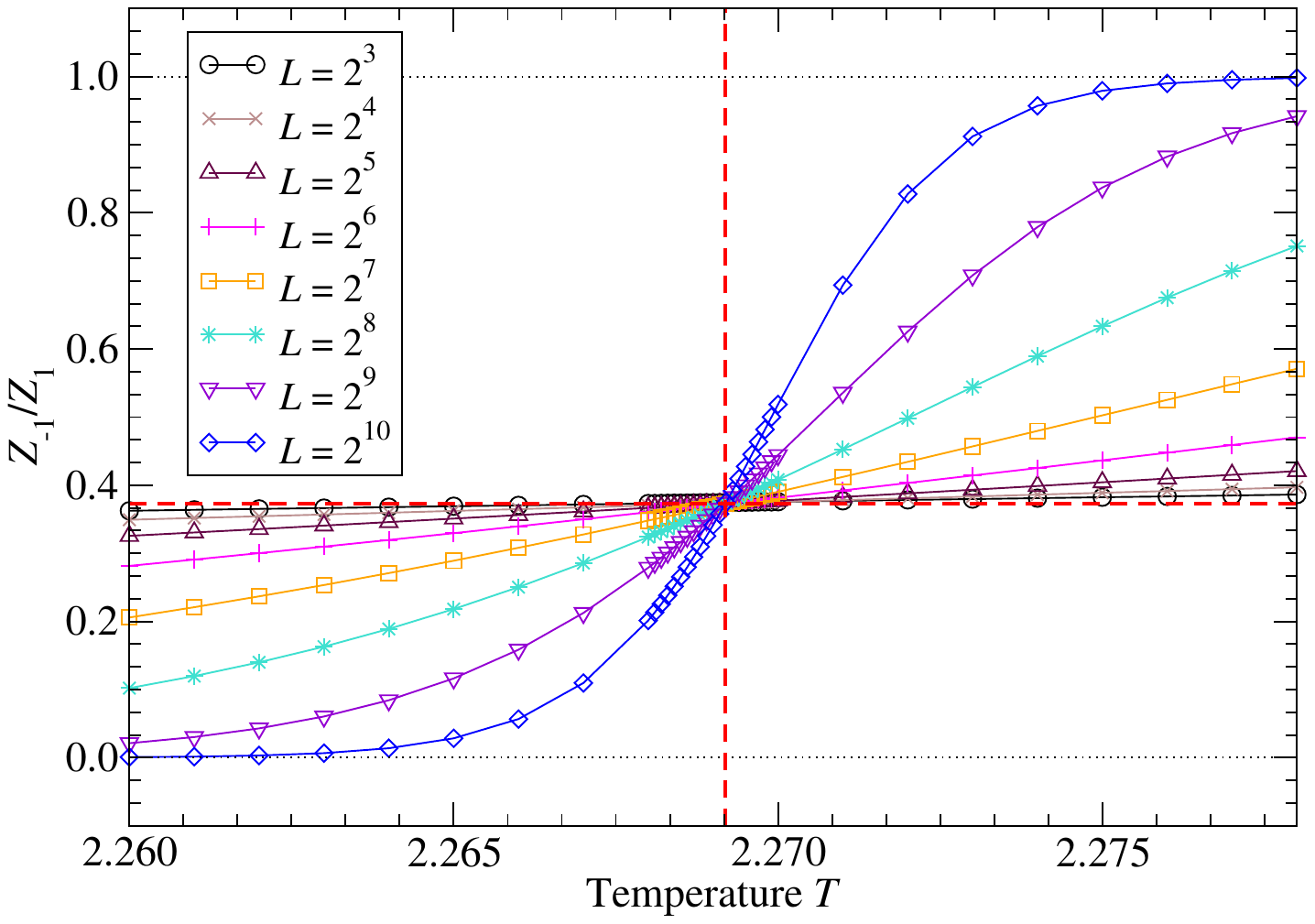}
    \caption{
        $Z_{-1}/Z_{1}$ in the 2D Ising model.
        The vertical dashed line indicates the exact critical point $T_{c}=2/(\log(1+\sqrt{2}))$.
        The horizontal dashed line denotes the exact value of $Z_{-1}/Z_{1}$ for the 2D Ising CFT, which is given by Eq.~\eqref{eq:exact_zpz_2DIsing}.
        The computation is done by the BTRG with the bond dimension $D_{\rm BTRG}=64$.
    }
    \label{fig:zpz_2DIsing}
\end{figure}

The partition function of the Ising model can be expressed as Eq.~\eqref{eq:part_TN} with the initial bond dimension 2, as
\begin{equation}
    (T_x)_{m_1m_2, m'_1m'_2}=\sqrt{I_{m_1}(\beta)I_{m_2}(\beta) I_{m'_1}(\beta)I_{m'_2}(\beta)}\, \delta_{m_1+m_2, m'_1+m'_2},
\end{equation} 
where $I_{0}(\beta)=\cosh\beta$ and $I_{1}(\beta)=\sinh\beta$, and we compute its tensor contraction using the BTRG with the bond dimension cutoff $D_{\rm BTRG}=64$.
Figure~\ref{fig:zpz_2DIsing} shows the ratio of the twisted partition functions $Z_{-1}(L\times L)/Z_{1}(L\times L)$ as a function of temperature for various volumes.
As the system size $L$ increases, the behavior of $Z_{-1}/Z_{1}$ exhibits a pronounced distinction across the critical point $T_{c}$: it approaches unity in the symmetric phase, while tending toward zero in the symmetry-broken phase, as expected.

We see that Figure~\ref{fig:zpz_2DIsing} also indicates that the curves of $Z_{-1}/Z_{1}$ with different volumes cross at a single point, 
implying the emergence of scale invariance, since the value of $Z_{-1}/Z_{1}$ at $T_c$ is independent of the system size. 
Thus, the ratio of the twisted partition functions serves the role of the Binder cumulant~\cite{Binder:1981sa} to determine $T_c$ with the TRG calculations. 
Moreover, its value at $T_c$ matches with the prediction of the $2$D Ising CFT~\cite{Ginsparg:1988ui} (see Appendix~\ref{appendix:CFT_twistedZ}): 
\begin{align}
    \label{eq:exact_zpz_2DIsing}
    \left.\frac{Z_{-1}(L_1\times L_2)}{Z_{1}(L_1\times L_2)}\right|_{T=T_c}
    = 
    \frac{|\theta_{3}(0,\tau)|+|\theta_{4}(0,\tau)|-|\theta_{2}(0,\tau)|}{|\theta_{3}(0,\tau)|+|\theta_{4}(0,\tau)|+|\theta_{2}(0,\tau)|},
\end{align}
where $\theta_{2}$, $\theta_{3}$, and $\theta_{4}$ denote Jacobi's theta functions, and $\tau={\rm i} L_2/L_1$ when we consider the model on the $L_1\times L_2$ square lattice. In Fig.~\ref{fig:zpz_2DIsing}, we set $\tau=\im$ as $L_1=L_2=L$ for the horizontal red dashed line. 
The scale-invariant point in the numerical results shows the nice consistency with the theoretical value of $Z_{-1}/Z_{1}$ in Eq.~\eqref{eq:exact_zpz_2DIsing}, which validates the efficacy of our approach.

\begin{figure}
    \centering
    \includegraphics[width=12cm]{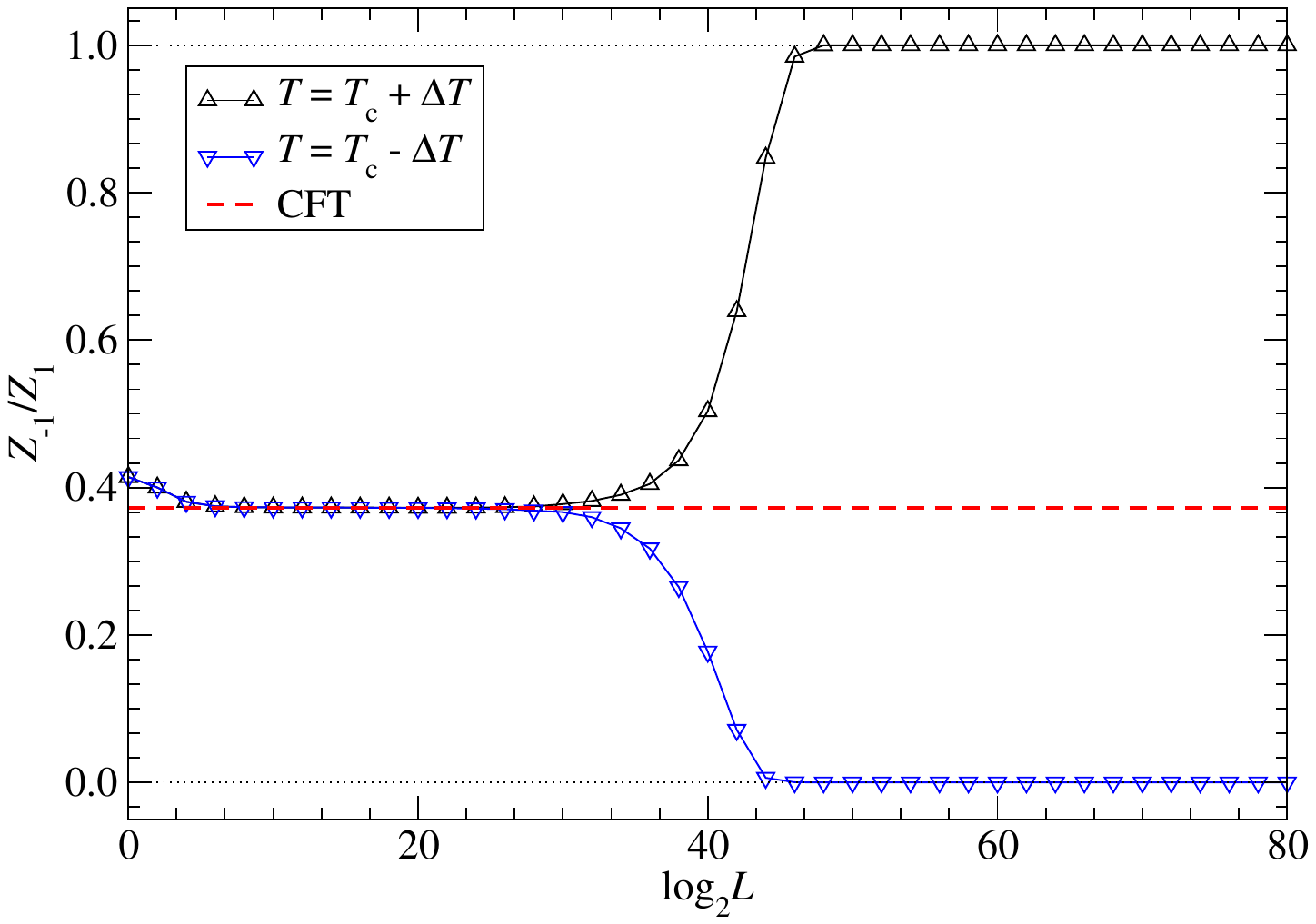}
    \caption{
        $Z_{-1}/Z_{1}$ as a function of $\log_{2}L$ in the 2D Ising model, with the temperature deviation $\Delta T=1.0\times10^{-6}$ from the critical point.
        The horizontal dashed line denotes the exact value of $Z_{-1}/Z_{1}$ for 2D Ising CFT.
        The computation is done by the BTRG with the bond dimension $D_{\rm BTRG}=64$.
    }
    \label{fig:zpz_diff_2DIsing}
\end{figure}

Figure~\ref{fig:zpz_diff_2DIsing} compares $Z_{-1}/Z_{1}$ in the vicinity of the critical point by setting $T=T_c\pm \Delta T$ with $\Delta T=1.0\times 10^{-6}$.
Even with such tiny deviations from the criticality, the twisted partition function successfully identifies whether the system is in the symmetric or symmetry-broken phase: as the system size increases, $Z_{-1}/Z_{1}$ converges to unity or zero, depending on whether the temperature lies in the symmetric or symmetry-broken phase, respectively.
This indicates that the ratio of the $\mathbb{Z}_{2}$-twisted partition function to the normal partition function allows one to bound the critical point in a manner analogous to the Gu--Wen ratio.

\begin{figure}
    \centering
    \includegraphics[width=12cm]{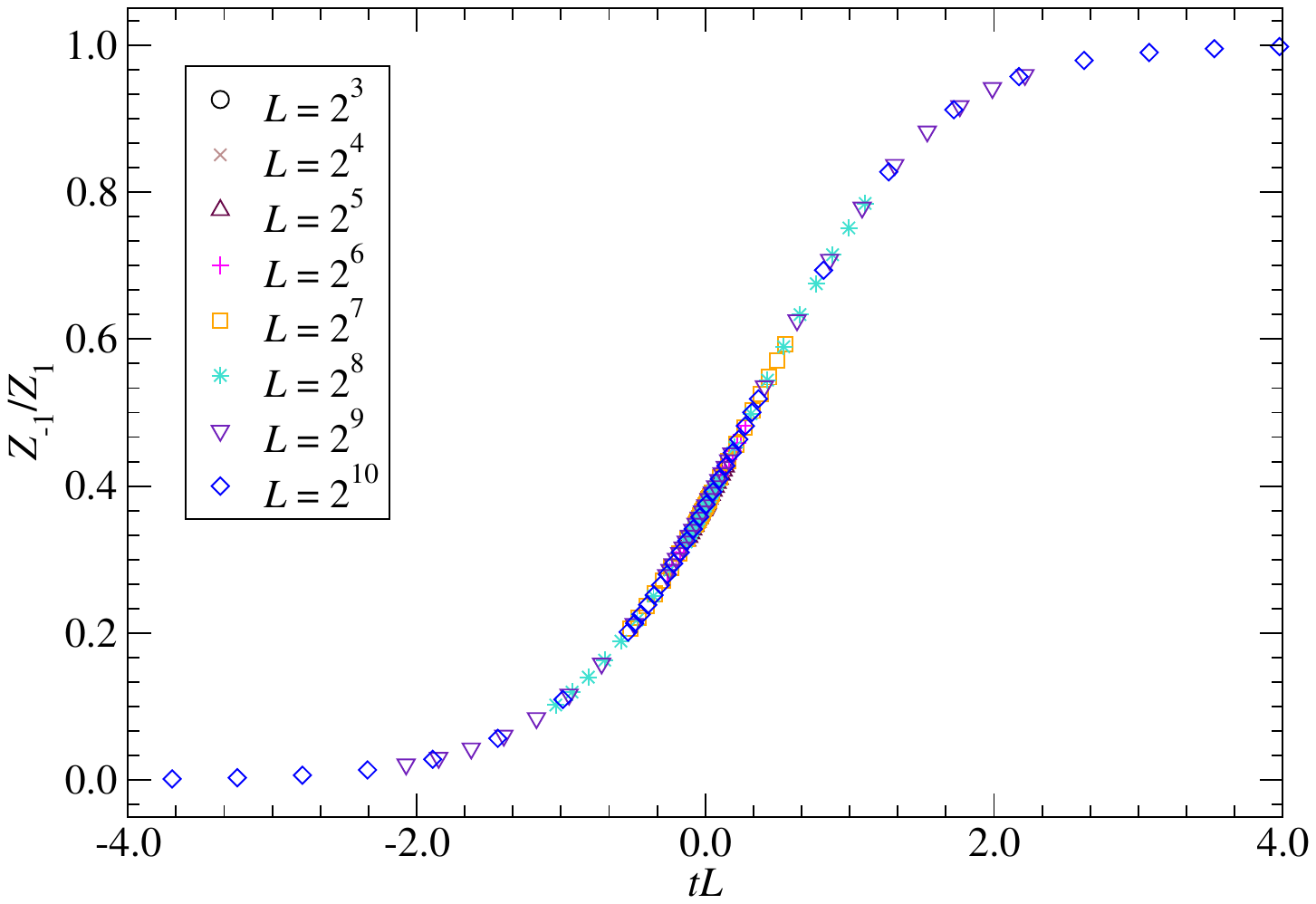}
    \caption{
        Finite-size scaling analysis for $Z_{-1}/Z_{1}$ as a function of $tL$ in the 2D Ising model. 
        The results of different volumes follow the universal curve as a function of $tL$ with $t=(T-T_c)/T_c$. 
        The computation is done by the BTRG with the bond dimension $D_{\rm BTRG}=64$.
    }
    \label{fig:zpz_FSS_2DIsing}
\end{figure}

We also confirm the finite-size scaling of $Z_{-1}/Z_{1}$ in the vicinity of the critical point. 
The 2D Ising CFT has three primary operators, the identity operator $1$, the spin operator $\sigma$, and the energy operator $\varepsilon$. 
The temperature $t=(T-T_{c})/T_{c}$ is the $\mathbb{Z}_2$-symmetric perturbation from the 2D Ising critical point, so it should correspond to the perturbation by the energy operator $\varepsilon$ that has the scaling dimension $\Delta=1$. 
Therefore, we can expect the universal curve as a function of $t L^{1/\nu}=tL$ since $\nu=1/(d-\Delta)=1$. 
Figure~\ref{fig:zpz_FSS_2DIsing} displays $Z_{-1}/Z_{1}$ as a function of $tL$, where the exact $T_{c}$ is employed for the definition of $t$. 
Data for different system sizes collapse onto a single curve when plotted against $tL$.
This demonstrates that $Z_{-1}/Z_{1}$ allows one to precisely determine both the critical point and the critical exponent, yielding $\nu=1$, while we here just substitute the known exact values for both $T_c$ and $\nu$.

Let us point out that the finite-size scaling analysis for the Gu--Wen ratio has been performed in Refs.~\cite{Morita:2024lwg}, and thus we would conclude that the ratio of the twisted partition functions and the Gu--Wen ratio equally work well for studying the critical phenomena associated with the discrete symmetry breaking.

\subsection{3D \texorpdfstring{$O(2)$}{O(2)} model}
\label{subsec:3DXY}

Next, let us consider an example for the critical phenomenon associated with the $U(1)$ symmetry breaking, and we study the 3D classical $O(2)$ model. 
The classical $O(2)$ model in $d$ dimensions is defined by
\begin{align}
    S=-\beta\sum_{x}\sum_{\mu=1}^{d}
    \cos(\theta_{x+\hat{\mu}}-\theta_{x}),
    \label{eq:classicalO2model}
\end{align}
where $\theta_n\in (-\pi,\pi]$ is the $U(1)$ variable.
The model can also be straightforwardly expressed in the form of Eq.~\eqref{eq:part_TN}, expanding ${\rm e}^{\beta\cos\Theta}$ in terms of the modified Bessel function of the first kind, $\rme^{\beta \cos \Theta}=\sum_{m=-\infty}^{\infty}I_m(\beta)\rme^{\im m \Theta}$, with setting the tensor components as 
\begin{equation}
\label{eq:tensor_O2}
    (T_x)_{m_1\ldots m_d, m'_1\ldots m'_d}=\sqrt{I_{m_1}(\beta)\cdots I_{m_d}(\beta) I_{m'_1}(\beta)\cdots I_{m'_d}(\beta)}\, \delta_{m_1+\cdots+m_d, m'_1+\cdots+m'_d}
\end{equation} 
at each lattice site $n$. The finite bond dimensional description of Eq.~\eqref{eq:tensor_O2} is obtained by truncating the character expansion, $\rme^{\beta \cos \Theta}\simeq\sum_{m=-K}^{K}I_m(\beta)\rme^{\im m \Theta}$, which results in the tensor network representation of the partition function with bond dimension $2K+1$~\cite{Liu:2013nsa}. In the range of $\beta$ considered in this study, taking $K\ge40$ already yields $I_K(\beta)/I_0(\beta)\lesssim 10^{-23}$, and the finite-$K$ effect is negligible.
We note that the finite bond-dimension cutoff necessarily suffers from the mild violation of the $U(1)$ symmetry, as it gives the upper and lower bounds for the $U(1)$ charge sectors, which violates the $U(1)$ conservation law: While the finite truncation preserves the global $U(1)$ invariance of the partition function, it violates the Ward-Takahashi identity when including the local operators. 
When we consider the twisted partition function, we take the twist angles quantized in $2\pi/N$ for some integer $N$, and thus we require that the truncations and the tensor decompositions are consistent with the charge conservation in mod $N$ by introducing the $N$ block structure. 

Let us consider the 3D $O(2)$ model in this subsection. 
For the twisted partition functions, we take the $\mathbb{Z}_2$ subgroup of the $U(1)$ symmetry, and the twist angle is set as $\alpha=\pi$. 
For the 3D $O(2)$ model, the continuous symmetry is expected to be spontaneously broken completely as $U(1)\to \{1\}$ at low temperatures, and the second-order phase transition takes place at some critical temperature $T_c$. In the low temperature limit, $\beta=1/T\gg 1$, we can take the quadratic approximation of the lattice action~\eqref{eq:classicalO2model} to obtain the continuum effective Lagrangian~\eqref{eq:effectiveLagrangian_U1SSB} of the $U(1)$ SSB phase with $v^2=\beta/a^{d-2}=\beta/a$ with $v$ as defined in ~\eqref{eq:effectiveLagrangian_U1SSB}, where $a$ is the lattice spacing. 
This suggests the $\mathbb{Z}_2$ twist does a sufficient job as the order parameter. 
The 3D $O(2)$ universality class has been extensively investigated in Monte Carlo simulations (see, for example, Refs.~\cite{Campostrini:2000iw,Komura_2014,Xu:2019mvy}). 
For the conformal bootstrap analysis about the 3D $O(2)$ criticality, see Refs.~\cite{Kos:2016ysd, Chester:2019ifh}. 
The application of the TRG approach to the $O(2)$ model at finite density has also been reported~\cite{Bloch:2021mjw}.

\begin{figure}
    \centering
    \includegraphics[width=12cm]{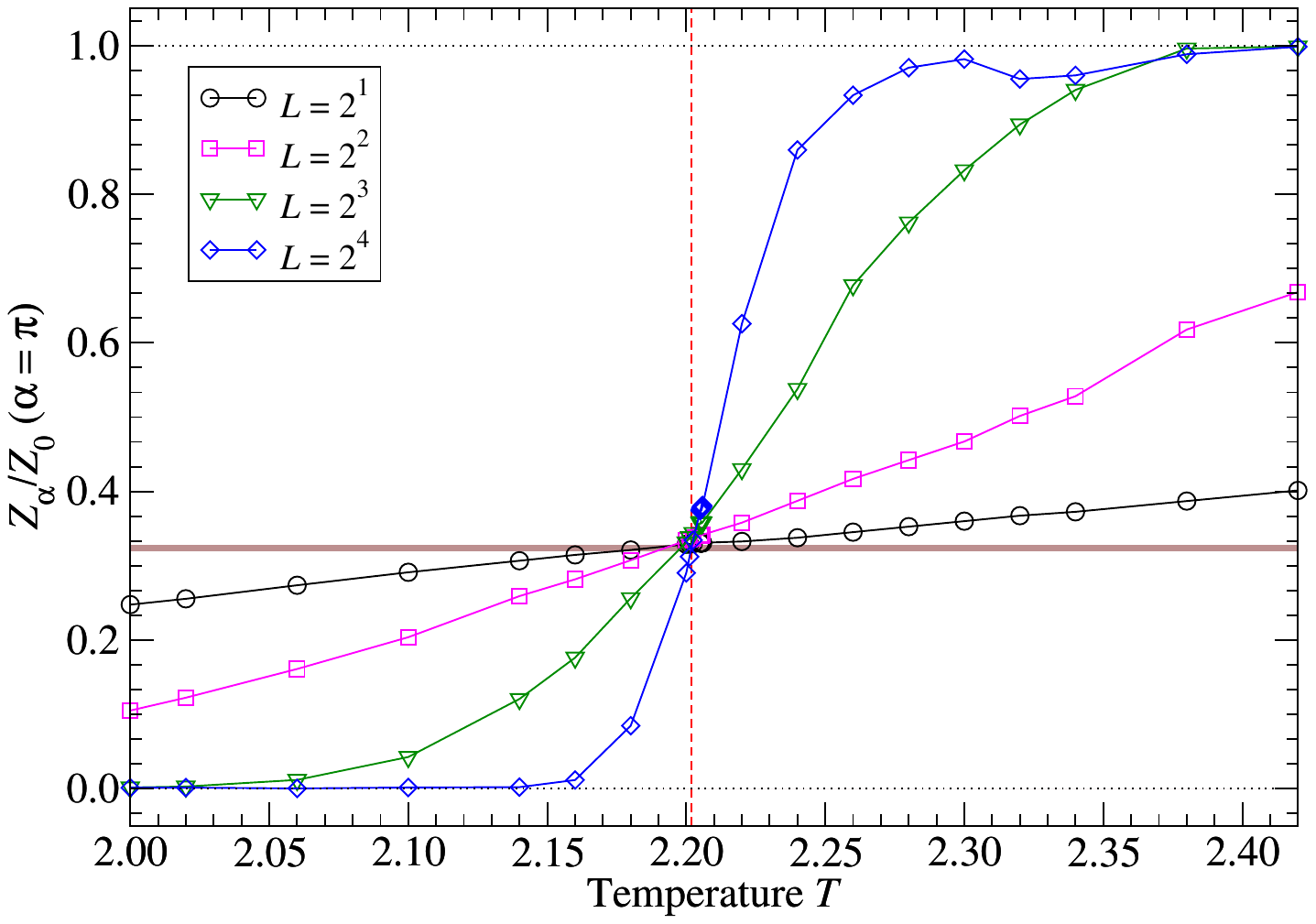}
    \caption{
        $Z_{\alpha=\pi}/Z_{0}$ in the 3D $O(2)$ model.
        The vertical dashed line indicates the critical point $T_c^{\mathrm{(MC)}} = 2.2018441(5)$ obtained by the Monte Carlo simulation in Ref.~\cite{Xu:2019mvy}. 
        The brown band denotes the value of $Z_{\alpha=\pi}/Z_{\alpha=0}$ at the critical point from the Monte Carlo simulation with $L=2^{4}$~\cite{Gottlob_1994}.
        The twisted partition function gives a clear signal for the $U(1)$ symmetry breaking. 
        The computation is done by the ATRG with the bond dimension $D_{\rm ATRG}=96$.
    }
    \label{fig:zpz_3DXY}
\end{figure}

Figure~\ref{fig:zpz_3DXY} presents the ratio of the twisted partition functions $Z_{\alpha=\pi}/Z_{\alpha=0}$ for various system sizes with $L\times L\times L$.\footnote{For the $U(1)$-twisted partition function, we write it as $Z_{\alpha}$ instead of $Z_{\rme^{\im \alpha}}$ for better readability. Thus, $Z_{\alpha=\pi}/Z_{\alpha=0}$ in the $O(2)$ model is exactly the same with $Z_{-1}/Z_{1}$ in the notation used for the Ising model. }
Here, we have focused on relatively smaller system sizes because the large volume results could be affected by artifacts from the finite bond dimension truncation~\cite{PhysRevB.108.024413},
and Appendix~\ref{appendix:3DXY} is devoted to discussing the finite bond-dimension effects, particularly in the thermodynamic limit.
The behavior of $Z_{\pi}/Z_0$ in Fig.~\ref{fig:zpz_3DXY} is qualitatively similar to that in Fig.~\ref{fig:zpz_2DIsing}; 
\begin{align}
    \frac{Z_{\alpha=\pi}(L\times L\times L)}{Z_{\alpha=0}(L\times L\times L)} 
    \xrightarrow{L\to \infty} \left\{
    \begin{array}{cl}
        1 &  (\text{in the symmetric phase $T>T_c$}),\\
        0 &  (\text{in the broken phase $T<T_c$}).
    \end{array}\right.
\end{align}
We can also see an emergent scale-invariant point at the phase transition $T=T_c$, which suggests that the phase transition is of the second order. 
The vertical red dotted line of Fig.~\ref{fig:zpz_3DXY} at $T_c^{(\mathrm{MC})}=2.2018441(5)$ is the critical temperature determined by the Monte Carlo simulation in Ref.~\cite{Xu:2019mvy}, which is consistent with our numerical result with the ATRG. 
In addition, the brown band in Fig.~\ref{fig:zpz_3DXY} denotes the resulting $Z_{\alpha=\pi}/Z_{\alpha=0}$ at the critical point from another Monte Carlo simulation with $L=2^{4}$~\cite{Gottlob_1994}, which again looks consistent with our result.

\begin{figure}
    \centering
    \includegraphics[width=12cm]{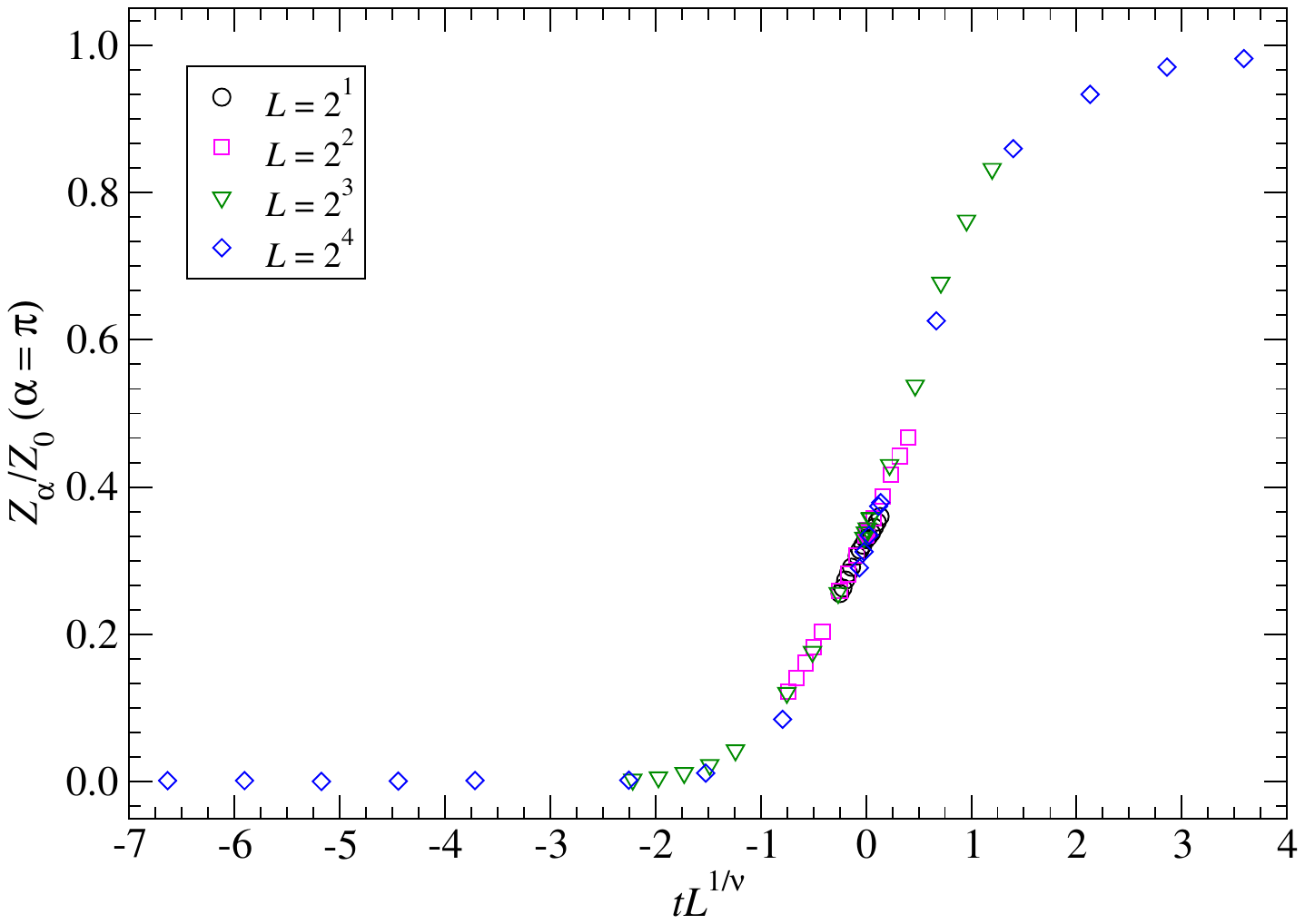}
    \caption{
        Finite-size scaling analysis for $Z_{\alpha=\pi}/Z_{0}$ in the 3D $O(2)$ model. 
        $Z_{\alpha=\pi}/Z_0$ with different volumes follow the universal curve when they are plotted as a function of $tL^{1/\nu}$. 
        The computation is done by the ATRG with the bond dimension $D_{\rm ATRG}=96$.
    }
    \label{fig:zpz_FSS_3DXY}
\end{figure}

To determine $T_c$, we first provide the upper and lower bounds for $T_{c}$ as $2.2012<T_c<2.2050$, 
which we set by observing the $L$-dependence of $Z_\pi/Z_0$ in Fig.~\ref{fig:zpz_3DXY} to judge whether the system spontaneously breaks $U(1)$ symmetry at a given temperature or not. 
For further precise determination of $T_{c}$, we employ the finite-size scaling of $Z_\pi/Z_0$ as we did in the 2D Ising model.
Introducing the reduced temperature, $t=(T-T_c)/T_c$, the results of different volumes should obey a universal curve in terms of $tL^{1/\nu}$ for some $\nu$. 
By changing candidate values of the critical temperature within the range $2.2012 < T_c < 2.2050$ in increments $\Delta T=0.0001$, we determine the optimal value of $\nu$ for each temperature by minimizing the deviation from a universal scaling function, without assuming any specific functional form, following the method of Ref.~\cite{doi:10.1143/JPSJ.62.435}.
Since the TRG results do not contain statistical errors, the cost function introduced in Ref.~\cite{doi:10.1143/JPSJ.62.435} can be interpreted as a direct measure of the deviation from the universal scaling curve. This deviation should vanish at the exact critical temperature and critical exponent for the data of sufficiently large volumes if the data points are fine enough after rescaling.
We find that the deviation takes its smallest value, approximately $6.7\times 10^{-3}$, when the assumed critical temperature lies in the range $2.2015\le T_{c}\le2.2019$. 
We therefore estimate the critical temperature as
\begin{equation}
    T_c=2.2017(2). 
\end{equation}
The scaling collapse is shown in Fig.~\ref{fig:zpz_FSS_3DXY}, and our result for $\nu$ is given by 
\begin{equation}
    \nu= 0.663(33),  
\end{equation} 
where the central value is determined by minimizing the cost function at $T_c=2.2017$, and the error bar is provided by allowing the $10\%$ deviation from the minimum of the cost function.
We note that the variation of $\nu$ obtained for changing the critical temperature as $2.2015\le T_{c}\le2.2019$ is found to be much smaller than this $10\%$ uncertainty, and we neglect it in our analysis.
Our computation of the critical exponent $\nu$ using the ratio of symmetry twisted and the standard partition function is the first precise estimate of the critical exponent for the 3D $O(2)$ model using the TRG approach. As reference values, the Monte Carlo and conformal bootstrap results for $\nu$ are $\nu^{(\mathrm{MC})}=0.67183(18)$~\cite{Xu:2019mvy} and $\nu^{(\mathrm{CB})}=0.67175(10)$~\cite{Chester:2019ifh}, respectively, and our TRG computation gives a consistent result, while the error is much larger than those of these techniques. 
We leave a more precise estimate for this exponent using the improved coarse-graining TRG algorithm~\cite{Lyu:2024lqh} or GPU-assisted tensor contractions~\cite{Jha:2023bpn} for future work.

\subsection{2D \texorpdfstring{$O(2)$}{O(2)} model}

Let us move on to the 2D classical $O(2)$ model, \eqref{eq:classicalO2model}. 
In contrast to dimensions three and higher, the Mermin--Wagner--Coleman theorem~\cite{Mermin:1966fe,Coleman:1973ci} prohibits the presence of the massless Nambu--Goldstone particle with the linear dispersion in two dimensions. 
Thus, the $U(1)$ symmetry does not get broken even at low temperatures, while the system is in the superfluid phase, showing the algebraic long-range order. 
The effective Lagrangian of this superfluid phase is given by \eqref{eq:effectiveLagrangian_U1SSB} when the vortex excitations are irrelevant. 

\begin{figure}
    \centering
    \subfigure[$\alpha=\pi$]{
    \includegraphics[width=8.6cm]{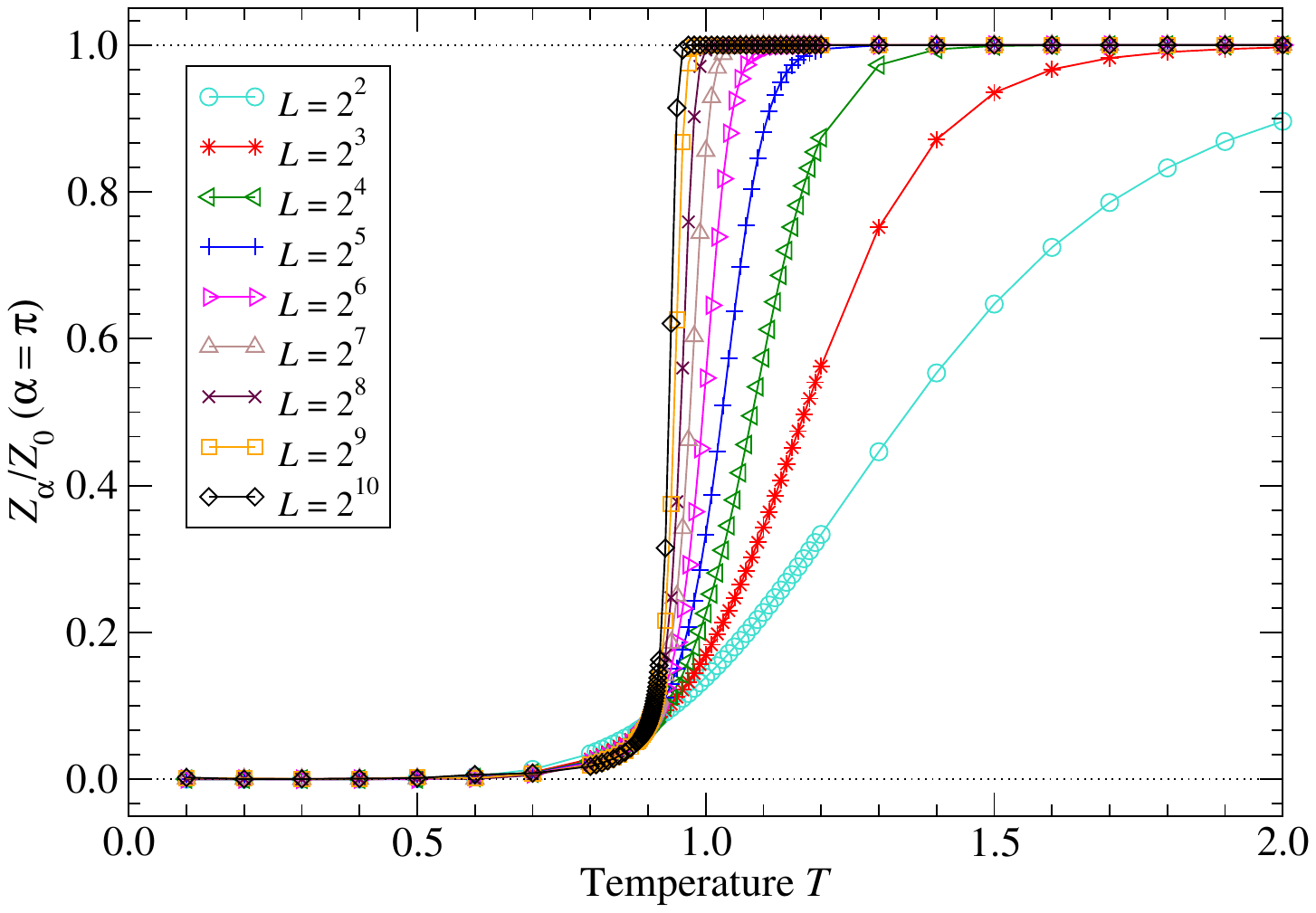}
    }
    \subfigure[$\alpha=\pi/2$]{
    \includegraphics[width=8.6cm]{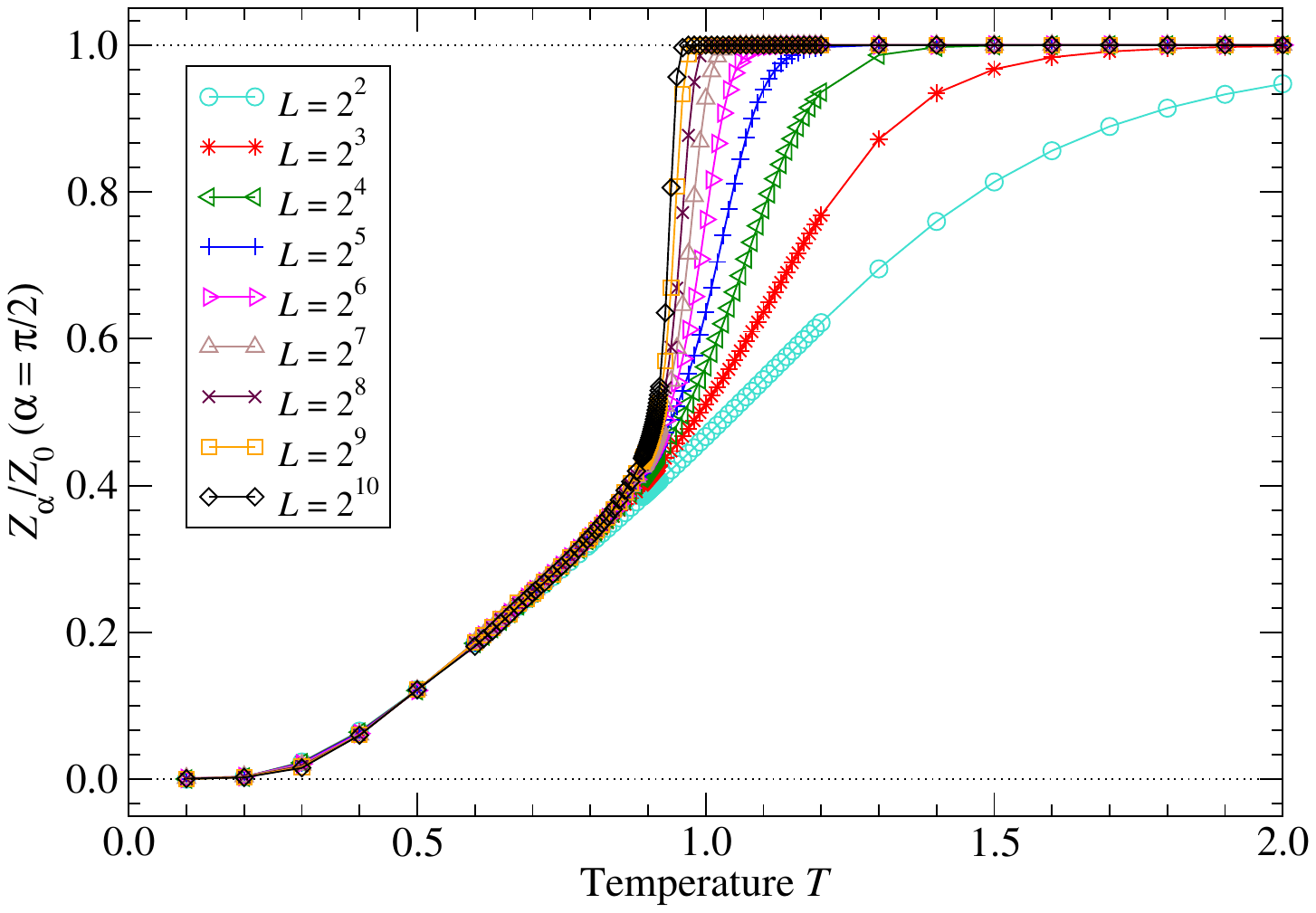}
    }
    \caption{
        $Z_{\alpha}(L\times L)/Z_{0}(L\times L)$ for the 2D $O(2)$ model with (a) $\alpha=\pi$, and (b) $\alpha=\pi/2$. 
        The result indicates that the system is in the $U(1)$-symmetric gapped phase at high temperatures, and it becomes a scale-invariant theory at low-temperatures with an exactly marginal deformation parametrized by $T$. 
        The computation is done by the BTRG with the bond dimension $D_{\rm BTRG}=200$.
    }
    \label{fig:zpz_zn_2DXY} 
\end{figure}

First, we compute $Z_{\alpha}(L\times L)/Z_0(L\times L)$ at twist angles, $\alpha=\pi$ and $\pi/2$, for the 2D $O(2)$ model, and the results are shown in Fig.~\ref{fig:zpz_zn_2DXY}. 
At high temperatures, we can observe the convergence $Z_\alpha/Z_0\rightarrow 1$ as $L\rightarrow \infty$, which suggests the unbroken $U(1)$ symmetry with the non-zero mass gap. 
At low temperatures, however, $Z_\alpha/Z_0$ does not converge to a definite value of unity or zero in the large-volume limit, which is in sharp contrast to the previous two models. 
Instead, they show scale-invariant behaviors at low temperatures, and thus the theory flows to conformal field theories with the exactly marginal parameter characterized by the temperature $T$. 
This observation is consistent with the BKT critical behavior.

To observe the BKT criticality in more detail, we evaluate the helicity modulus $\Upsilon_\alpha$ using the twisted partition function $Z_{\alpha}/Z_0$ at finite twist angles:
\begin{align}
    \Upsilon_\alpha(L_1\times L_2)
    =
    -T\frac{2}{\alpha^2}\frac{L_2}{L_1}\log\frac{Z_\alpha(L_1\times L_2)}{Z_0 (L_1\times L_2)},
\end{align}
where the twist is imposed along the $L_2$-direction, and the $L_1$-direction is periodic. 
In the limit $\alpha\to 0$, the definition matches with the helicity modulus computed from the local correlation function. 
At high temperatures ($T>T_{\mathrm{BKT}}$), $\Upsilon_{\alpha}\to 0$ in the large-volume limit. 
On the superfluid regime ($T<T_{\mathrm{BKT}}$), $\Upsilon_\alpha$ remains to be finite, and the theoretical computation~\eqref{eq:twistedZ_BKTsuperfluid} suggests 
\begin{equation}
    \Upsilon_\alpha(L_1\times L_2)=\frac{T}{2\pi R_{\mathrm{ren}}^2} + O(\rme^{-\# \frac{L_1}{L_2}})
    \label{eq:HelicityModulus_smallAlpha}
\end{equation}
for $0<\alpha<\pi$, where $R_{\mathrm{ren}}$ is the renormalized radius of the dual compact boson.\footnote{As the quadratic approximation of the Lagrangian gives $(2\pi R^2)^{-1}=\beta$ for the dual-compact-boson radius in $d=2$, one can expect $T/(2\pi R_{\mathrm{ren}}^2) \to 1$ in the zero-temperature limit ($T=0$), where the fluctuation effects are suppressed. } 
However, the behavior at $\alpha=\pi$ has an extra contribution as there are two equal-length paths connecting antipodal points,
\begin{equation}
    \Upsilon_\pi(L_1\times L_2)=\frac{T}{2\pi R_{\mathrm{ren}}^2}-\frac{2T}{\pi^2}\frac{L_2}{L_1}\ln 2 + O(\rme^{-\# \frac{L_1}{L_2}}). 
    \label{eq:HelicityModulus_AlphaPi}
\end{equation}
In other words, one of the higher-winding contributions in $O(\rme^{-{\# L_1/L_2}})$ becomes comparable with the leading contribution as $\alpha \to \pi$. 

Equation~\eqref{eq:HelicityModulus_smallAlpha} shows that the helicity modulus $\Upsilon_\alpha$ dictates the renormalized compact boson radius directly for $0<\alpha\ll \pi$ when we can neglect the higher-winding contributions of $O(\rme^{-\# L_1/L_2})$. This approximation becomes valid when $L_1/L_2 \gg 1$.\footnote{Ideally, it would be desired to take the limit $L_1/L_2\to \infty$ for completely eliminating higher-winding contributions. However, we should note that we compute $\Upsilon_\alpha$ out of $Z_{\alpha}/Z_{0}$, and this partition-function ratio itself becomes exponentially small as $L_1/L_2\to \infty$ according to \eqref{eq:twistedZ_BKTsuperfluid}. Thus, we should vary the ratio $L_1/L_2$ within the regime where $Z_\alpha/Z_0$ is computable with good accuracy. Moreover, as the higher-winding contributions are exponentially suppressed, we do not need to take extremely large $L_1/L_2$ in practice.} 
This relationship becomes important when we apply the Nelson--Kosterlitz criterion~\cite{Nelson:1977zz} to obtain the BKT transition temperature $T_{\mathrm{BKT}}$, because it compares the helicity modulus $\Upsilon_\alpha$ with $2T/\pi$ to identify the marginality crossing point $R_{\mathrm{ren}}^2=1/4$. 
We note that this aspect-ratio dependence exists also for the helicity modulus defined by the local correlation function (or $\Upsilon_\alpha$ with the $\alpha\to 0$ limit), and see Refs.~\cite{PhysRevB.61.11282, PhysRevB.69.014509}, which study the aspect-ratio dependence of the helicity modulus. 
Although these higher-winding contributions turn out to be negligibly small within the numerical accuracy of this paper,\footnote{For the $L\times L$ lattice, its contribution is already $O(10^{-4})$, and it becomes $O(10^{-9})$ for the $2L\times L$ lattice. }  we perform the computation not only at the $L\times L$ lattice but also at the $2L\times L$ lattice to suppress them.

\begin{figure}
    \centering
    \includegraphics[width=12cm]{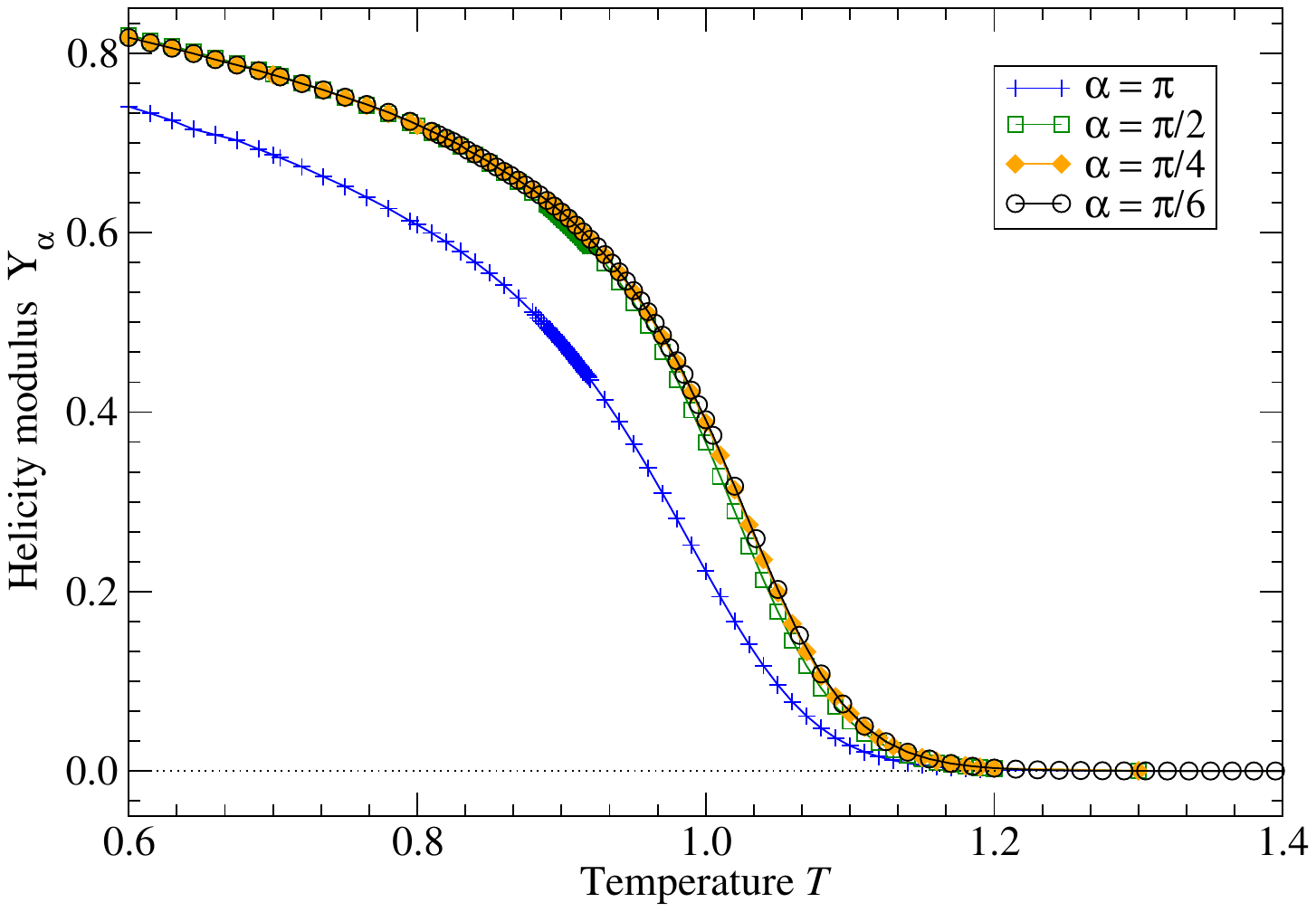}
    \caption{
        Helicity modulus $\Upsilon_\alpha$ at several twist angles $\alpha$ in the 2D $O(2)$ model on an $L\times L$ lattice with $L=32$. 
        The results at low temperatures fit with the theoretical expectation based on the compact-boson CFT: The $\alpha$-dependence is quite tiny for $0<\alpha\ll \pi$, and $\Upsilon_{\alpha=\pi}$ is significantly smaller than the others. 
        The computation is done by the BTRG with the bond dimension $D_{\rm BTRG}=200$.
    }
    \label{fig:helicity_R1_iter10}
\end{figure}

Figure~\ref{fig:helicity_R1_iter10} shows the helicity modulus $\Upsilon_\alpha(L\times L)$ evaluated at several twist angles, $\alpha=\pi, \pi/2, \pi/4, \pi/6$, for fixed $L=32$.
We observe that the twist-angle dependence is quite tiny for $\Upsilon_{\alpha=\pi/2,\pi/4,\pi/6}$, which is consistent with the theoretical formula~\eqref{eq:HelicityModulus_smallAlpha} as its $\alpha$-dependence appears only in the subleading higher-winding contributions $O(\rme^{-\# L_1/L_2})$. 
Furthermore, $\Upsilon_{\alpha=\pi}$ is significantly lower than the others on the low-temperature regions, which also matches with the theoretical expectation~\eqref{eq:HelicityModulus_AlphaPi}. 
These helicity moduli at finite $\alpha$, obtained directly from the twisted partition functions, are found to be consistent with the recent Monte Carlo simulation results in Ref.~\cite{Khairnar_2025}.

 \begin{figure}
    \centering
    \subfigure[$L\times L$ lattice]{
    \includegraphics[width=8.6cm]{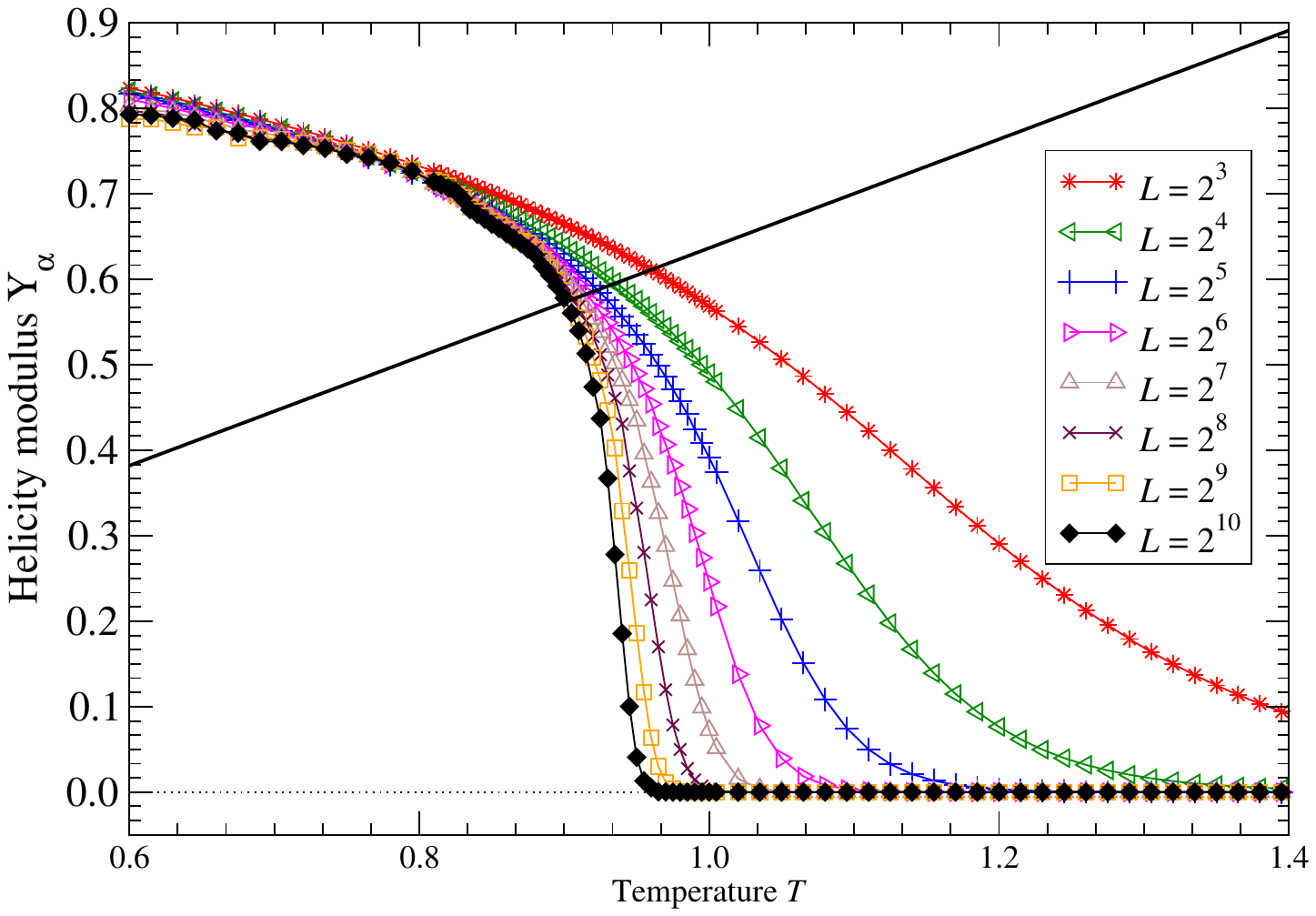}
    }
    \subfigure[$2L\times L$ lattice]{
    \includegraphics[width=8.6cm]{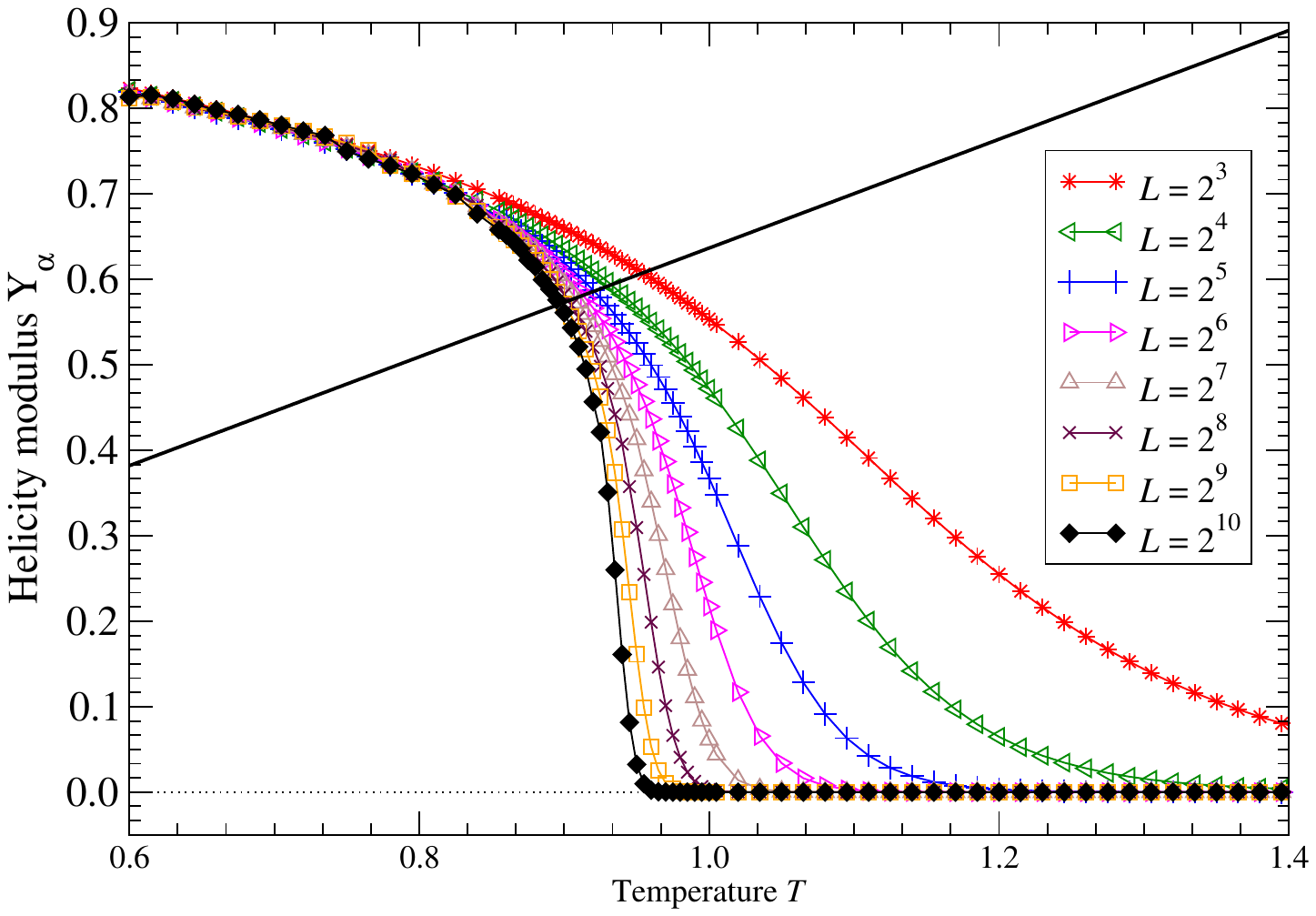}
    }
    \caption{
        Helicity modulus at $\alpha=\pi/6$ for varying system sizes at fixed aspect ratios: (a) $L_1/L_2=1$, (b) $L_1/L_2=2$.
        The solid line denotes the Nelson--Kosterlitz formula, $\Upsilon_\alpha=2T/\pi$~\cite{Nelson:1977zz}. 
        It turns out that the results for the $2L\times L$ lattice is numerically more stable than the ones for the $L\times L$ lattice. 
        The computation is done by the BTRG with the bond dimension $D_{\rm BTRG}=200$.
    }
    \label{fig:HelicityModulus_2DXY} 
\end{figure}

Figure~\ref{fig:HelicityModulus_2DXY} shows the helicity modulus $\Upsilon_{\alpha=\pi/6}$ for the $L\times L$ and $2L\times L$ lattices with various volumes. 
On the high-temperature sides near $T_{\mathrm{BKT}}$, we observe the so-called ``universal jump'' of the helicity modulus as originally predicted in Ref.~\cite{Nelson:1977zz}.
We find that the numerical results of the $2L\times L$ lattice are more stable than the ones of the $L\times L$ lattice: 
Since the low-temperature superfluid phase flows to a CFT fixed point, $\Upsilon_\alpha$ is expected to become $L$-independent for sufficiently large $L$, which is achieved well for the $2L\times L$ lattice but not satisfactorily for the $L\times L$ lattice.  
This phenomenon may be attributed to the observation that the effective bond dimension of the $2L\times L$ lattice becomes much larger than that of the $L\times L$ lattice~\cite{Yang:2017lvo}. Therefore, we mainly use the results of the $2L\times L$ lattice for the determination of $T_{\mathrm{BKT}}$.

To determine $T_{\mathrm{BKT}}$, we first determine the temperature $T^*(L)$, at which the helicity modulus $\Upsilon_{\alpha}(2L\times L)$ crosses with the Nelson--Kosterlitz formula, and we take its extrapolation as $L\to \infty$ assuming the following scaling relation,
\begin{align}
\label{eq:fitting_ansatz}
    T^{*}(L)
    =
    T_{\rm BKT}
    +
    \frac{a}{(\ln bL)^{2}},
\end{align}
where $a$ and $b$ are non-universal fitting parameters.
This is one of the standard procedures to determine the BKT transition temperature (see, e.g., Ref.~\cite{PhysRevB.69.014509,Khairnar_2025,Sandvik:2010lkj}) and we follow this conventional method, but we note that Ref.~\cite{Hsieh_2013} takes into account the sub-leading logarithmic correction for the helicity modulus in the determination of $T^*(L)$, which results in the higher BKT transition temperature. 
Let us leave it for a future study to investigate if such a higher-order logarithmic correction for the Nelson--Kosterlitz criterion is actually present using the TRG method.

 \begin{figure}
    \centering
    \subfigure[$\alpha=\pi$]{
    \includegraphics[width=8.6cm]{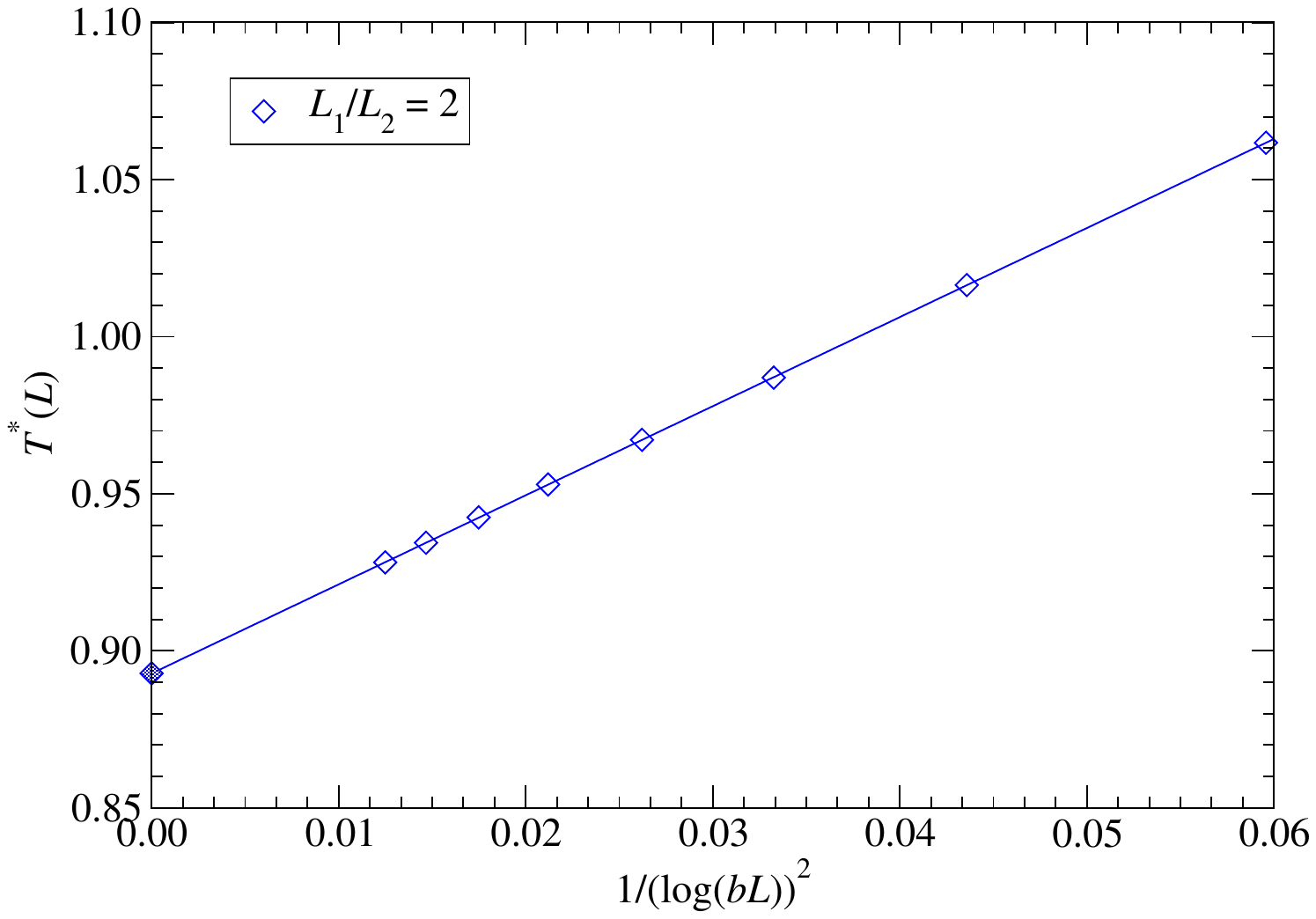}
    }
    \subfigure[$\alpha=\pi/6$]{
    \includegraphics[width=8.6cm]{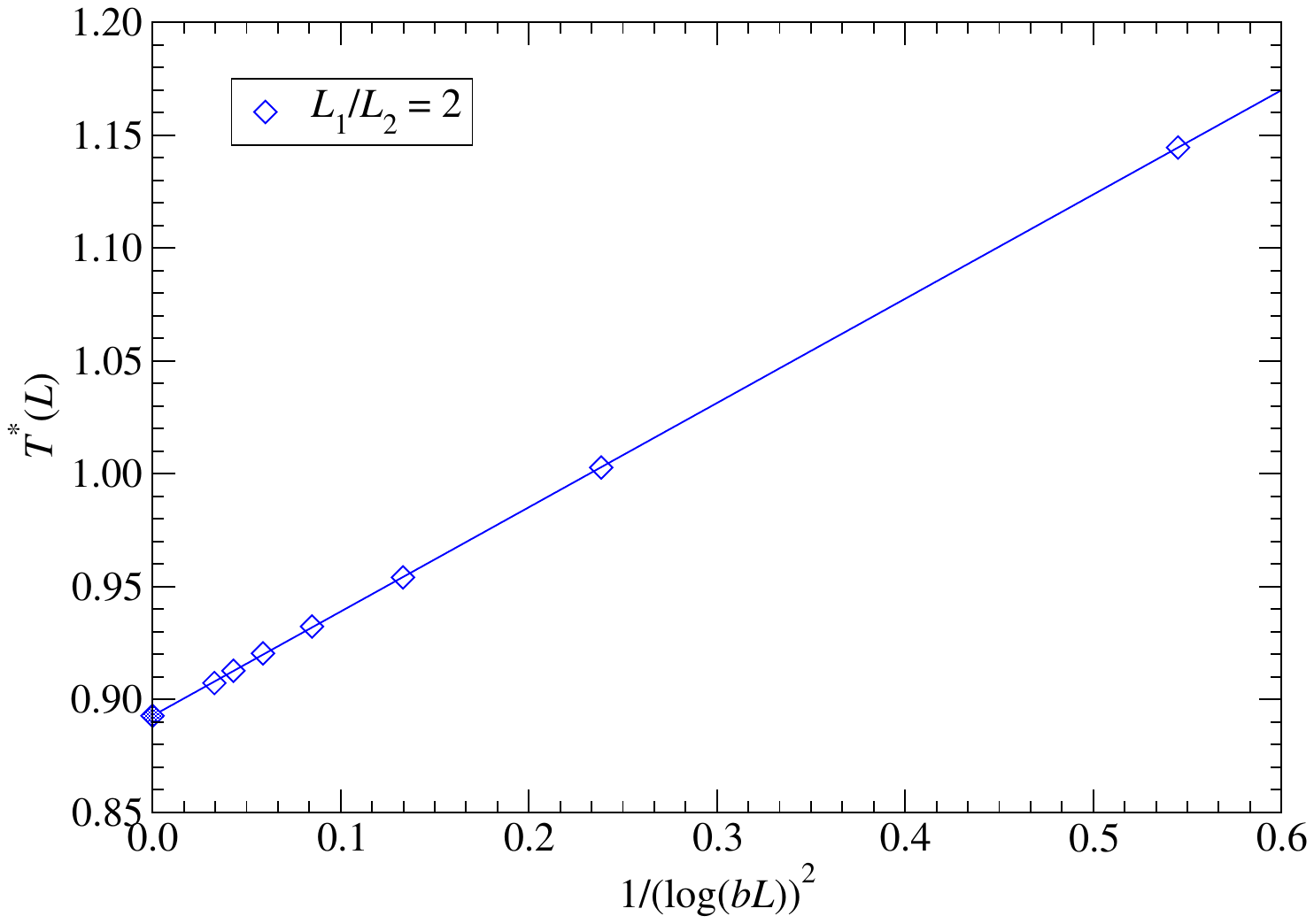}
    }
    \caption{
        $T^{*}(L)$ as a function of $1/(\ln bL)^{2}$ for varying system sizes with the $2L\times L$ lattices: (a) $\alpha=\pi$, (b) $\alpha=\pi/6$. 
        For $\alpha=\pi$, we determined $T^*(L)$ by using the modified Nelson--Kosterlitz criterion based on \eqref{eq:HelicityModulus_AlphaPi}. 
        The results of different volumes follow the universal curve as a function of $1/(\ln b L)^2$, and we extract the BKT transition temperature by extrapolating $L\to \infty$. 
    }
    \label{fig:estimated_Tc} 
\end{figure}

Figure~\ref{fig:estimated_Tc} shows $T^{*}(L)$ against $1/(\ln bL)^{2}$ at $\alpha=\pi$ and $\alpha=\pi/6$ evaluated on the $2L\times L$ lattices. 
When $\alpha=\pi$, $T^{*}(L)$ is determined based on Eq.~\eqref{eq:HelicityModulus_AlphaPi}, not on the standard Nelson--Kosterlitz formula~\eqref{eq:HelicityModulus_smallAlpha}, while the standard one is used for $\alpha=\pi/6$. 
In both cases, we simply neglect the correction from the higher-winding contributions as they are sufficiently small for the $2L\times L$ lattice with the current precision. 
The result is 
\begin{align}
    T_{\mathrm{BKT}}=\left\{\begin{array}{cl}
       0.8928(2)  &  (\text{from the $\alpha=\pi$ TRG}),\\
       0.8927(5)  &  (\text{from the $\alpha=\pi/6$ TRG}).
    \end{array}
    \right.
\end{align}
Both results are consistent not only with each other but also with previous literature~\cite{Hasenbusch:2005xm,Vanderstraeten:2019frg,Jha:2020oik,PhysRevB.104.165132}.

\section{Conclusions and Discussions}
\label{sec:Conclusion}

The symmetry-twisted partition function provides an order parameter to detect phase transitions in spin models with either discrete or continuous global symmetries, and we have demonstrated that the TRG method is an excellent framework for its computation. 
At the second-order phase transition point, finite-size scaling theory works in the same way as the case of the Binder cumulant, but computing the symmetry-twisted partition function is much easier for the TRG method because the local correlation functions are totally unnecessary. 
For its explicit demonstration, we analyze the 2D classical Ising model and the 3D classical $O(2)$ model, and their critical temperature $T_c$ and the scaling exponent $\nu$ are successfully obtained. 
In the case of the 3D $O(2)$ model, this gives the first accurate TRG computation of $\nu$ to our knowledge. 
We also study the 2D $O(2)$ model and observe the BKT critical behavior at low temperatures, and we determine the BKT transition temperature using the Nelson--Kosterlitz criterion for the helicity modulus. 
It is an important future study if the TRG computation of the symmetry-twisted partition function becomes competitive with other conventional frameworks, such as the Monte Carlo method, for these quantities.

One of the potentially interesting applications is to consider the generalized $O(2)$ model, which extends the lattice action~\eqref{eq:classicalO2model} of the usual $O(2)$ model as~\cite{osti_5670624, PhysRevLett.55.541}
\begin{equation}
    S=-\sum_{x,\mu}\left\{\beta_1\cos(\theta_{x+\hat{\mu}}-\theta_x)+\beta_q \cos(q(\theta_{x+\hat{\mu}}-\theta_x))\right\}. 
\end{equation}
For the TRG study of this model with $d=2$, see Ref.~\cite{Samlodia:2024kyi}. For $d\ge 3$, if the phase transition is dominated by the $\beta_1$ term, the SSB pattern is given by $U(1)\xrightarrow{\mathrm{SSB}} \{1\}$, while if it is caused by the $\beta_q$ term, we expect $U(1)\xrightarrow{\mathrm{SSB}}\mathbb{Z}_q$. 
The symmetry-twisted partition function can clearly discriminate these SSB patterns by changing the twist angles: When $\alpha\in \frac{2\pi}{q}\mathbb{Z}$, the twisted partition function is only sensitive for $U(1)\xrightarrow{\mathrm{SSB}}\{1\}$, while it becomes sensitive for both SSB patterns when $\alpha\not\in \frac{2\pi}{q}\mathbb{Z}$. 
Combining these different twists can help uncover the complete SSB pattern of this model in the $\beta_1$-$\beta_q$ phase diagram.

It would also be interesting to apply the idea of twisted boundary conditions to the study of gauge theories using TRG. 
As a nontrivial example of such a direction, we can study a model of quantum gauge theories that becomes a topologically ordered phase in the infrared and supports nontrivial anyons for its low-energy excitations. 
Even when the UV theory only has a $0$-form global symmetry, those anyons often carry the projective representations, a phenomenon which is called the symmetry fractionalization~\cite{Chen:2014wse, Barkeshli:2014cna, Hsin:2019fhf, Delmastro:2022pfo}: The background gauge field for the emergent higher-form symmetry is activated via the background gauge field for the $0$-form symmetry. 
When this happens, the twisted partition function enjoys a characteristic behavior that is different from those of SSB phases, and thus the symmetry-twisted partition function becomes an order parameter for the topologically ordered phase with the nontrivial symmetry fractionalization~\cite{Maeda:2025ycr}.

\begin{acknowledgments}

We are grateful to Satoshi Morita, Shinsuke Nishigaki, Tomotoshi Nishino, Tsuyoshi Okubo, Hidemaro Suwa, and Synge Todo for their valuable comments.
The numerical computation for the present work was carried out with SQUID at the Cybermedia Center, Osaka University (Project ID: hp250055, G16353), and with Pegasus provided by the Multidisciplinary Cooperative Research Program of Center for Computational Sciences, University of Tsukuba.

S. A. acknowledges the support from JSPS KAKENHI Grant Numbers JP23K13096 and JP25H01510, the Center of Innovations for Sustainable Quantum AI (JST Grant Number JPMJPF2221), the Endowed Project for Quantum Software Research and Education, the University of Tokyo~\cite{qsw}, and the Top Runners in Strategy of Transborder Advanced Researches (TRiSTAR) program conducted as the Strategic Professional Development Program for Young Researchers by the MEXT. 
R.~G.~J is supported by the U.S. Department of Energy, Office of Science, Advanced Scientific Computing Research, under contract number DE-SC0025384. 
J. M. is supported by JST SPRING, Grant Number JPMJSP2110.
The work of Y.~T. is partially supported by JSPS KAKENHI Grant No. 23K22489 and also by Center for Gravitational Physics and Quantum Information (CGPQI) at Yukawa Institute for Theoretical Physics (YITP). 
This manuscript has been co-authored by an employee of Fermi Research Alliance, LLC under Contract No. DE-AC02-07CH11359 with the U.S. Department of Energy, Office of Science, Office of High Energy Physics.  This work is supported by the Department of Energy through the Fermilab Theory QuantiSED program in the area of ``Intersections of QIS and Theoretical Particle Physics''.

Lastly, discussions during the long-term workshop, HHIQCD2024, at YITP (YITP-T-24-02), were valuable for this work.
We are also benefited from discussions at the YITP workshop (YITP-I-25-02) on ``Recent Developments and Challenges in Tensor Networks: Algorithms, Applications to Science, and Rigorous Theories."
\end{acknowledgments} 

\section*{Data availability}
The data that support the findings of this article are openly available at Ref.~\cite{akiyama_2026_18169605}.

\appendix

\section{Twisted partition functions of 2D CFT}
\label{appendix:CFT_twistedZ}

We briefly review the discussion of the twisted partition function in 2D CFT~\cite{Petkova:2000ip}. The torus partition function of 2D CFT with the modular parameter $\tau$ is given by
\begin{align}
    Z(\tau) = \Tr_{\mathcal{H}}\left( q^{L_0-c/24} \bar{q}^{\bar{L}_0-c/24} \right),
\end{align}
where $q=\e^{2\pi\im\tau}$ and $c$ is the central charge. Since the Hilbert space $\mathcal{H}$ can be decomposed into the direct sum of left- and right-moving Virasoro representations, the torus partition function can be expressed as
\begin{align}
    Z(\tau) = \sum_{h,\bar{h}}M_{h,\bar{h}}\chi_h(\tau)\bar{\chi}_{\bar{h}}(\bar{\tau}),
\end{align}
where $M_{h,\bar{h}}$ denotes the multiplicity of the representation with highest weight $(h,\bar{h})$ in $\mathcal{H}$ and
\begin{align}
    \chi_h(\tau) = \Tr_{V_h}\left(q^{L_0-c/24} \right)
\end{align}
is the Virasoro character.

The twisted torus partition function is defined as
\begin{align}
    Z_U(\tau) = \Tr_{\mathcal{H}}\left(U q^{L_0-c/24}\bar{q}^{\bar{L}_0-c/24}\right),
\end{align}
where $U$ is a symmetry transformation operator, which does not have to be invertible~\cite{Chang:2018iay}. 
%\footnote{$U$ is not necessarily invertible.}. 
Since $U$ commutes with the Virasoro generators, the twisted partition function $Z_U(\tau)$ can be computed solely from the action of $U$ on the primary states. 
Below, we discuss two well-known examples.

The first example is a diagonal rational CFT (RCFT). In a diagonal RCFT, there are symmetry operators, which have one-to-one correspondence with primary operators. These symmetry operators are so-called Verlinde lines, and their actions on primary states were studied in Ref.~\cite{Verlinde:1988sn}. We denote a primary state and the corresponding Verlinde line by $\ket{\phi_i}$ and $\mathcal{L}_i$, respectively. Then, the action of $\mathcal{L}_i$ on $\ket{\phi_j}$ is given by
\begin{align}
    \mathcal{L}_i\ket{\phi_j} = \frac{S_{ij}}{S_{0j}}\ket{\phi_j},
\end{align}
where $S_{ij}$ is the modular $S$-matrix. Then, the torus partition function with an insertion of $\mathcal{L}_i$ is given by
\begin{align}
    Z_{U=\mathcal{L}_i}(\tau) = \sum_{j}\frac{S_{ij}}{S_{0j}}|\chi_j(\tau)|^2. 
\end{align}

For instance, the Ising CFT has three primaries $1,\varepsilon,\sigma$ with conformal weights $(0,0), (\frac{1}{2},\frac{1}{2}), (\frac{1}{16},\frac{1}{16})$, respectively. The Verlinde line corresponding to $\varepsilon$ is the $\mathbb{Z}_2$ symmetry line $\eta$ of the spin flip in the Ising model. Since $1,\varepsilon$ is even, while $\sigma$ is odd under this $\mathbb{Z}_2$ line, the twisted partition function is given by
\begin{align}
    Z_\eta(\tau) = |\chi_0|^2 + |\chi_{\frac{1}{2}}|^2 - |\chi_{\frac{1}{16}}|^2.
\end{align}
Taking the ratio of the twisted and untwisted partition function, we obtain
\begin{equation}
    \frac{Z_\eta(\tau)}{Z(\tau)} = \frac{|\chi_0|^2 + |\chi_{\frac{1}{2}}|^2 - |\chi_{\frac{1}{16}}|^2}{|\chi_0|^2 + |\chi_{\frac{1}{2}}|^2 + |\chi_{\frac{1}{16}}|^2} = \frac{|\theta_3(\tau)|+|\theta_4(\tau)|-|\theta_2(\tau)|}{|\theta_3(\tau)|+|\theta_4(\tau)|+|\theta_2(\tau)|},
\end{equation}
where $\theta_i(\tau)$ denote the Jacobi theta functions. In the notation of Sec.~\ref{sec:NumericalResults_2DIsing}, $Z_\eta(\tau)=Z_{-1}(L_1\times L_2)$ and $Z(\tau)=Z_{1}(L_1\times L_2)$ (up to the gravitational local counter term) with $\tau=\im L_2/L_1$ when $T=T_c$. 

The second example is the compact boson CFT with the radius $R$. A primary operator of the compact boson CFT is labeled by a momentum charge $n$ and a winding charge $w$, and it has conformal weight
\begin{align}
    h = \frac{1}{4}\left(\frac{n}{R}+wR\right)^2,
    \qquad \bar{h} = \frac{1}{4}\left(\frac{n}{R}-wR\right)^2.
\end{align}
Therefore, the untwisted torus partition function is obtained by
\begin{align}
    Z(\tau) = \frac{1}{|\eta(\tau)|^2} \sum_{n,w\in\mathbb{Z}} q^{\frac{1}{4}\left(\frac{n}{R}+wR\right)^2}\bar{q}^{\frac{1}{4}\left(\frac{n}{R}-wR\right)^2}.
\end{align}
The compact boson CFT has $U(1)$ momentum symmetry, which we denote a symmetry operator by $U_{\alpha}$. Since a primary operator labeled by $(n,w)$ has charge $n$ under this symmetry, the twisted partition function is obtained by
\begin{align}
    Z_{\alpha}(\tau) = \frac{1}{|\eta(\tau)|^2} \sum_{n,w\in\mathbb{Z}} \e^{\im\alpha n}q^{\frac{1}{4}\left(\frac{n}{R}+wR\right)^2}\bar{q}^{\frac{1}{4}\left(\frac{n}{R}-wR\right)^2}.
\end{align}
From this expression with $\tau=\im L_2/L_1$, one can easily check~\eqref{eq:twistedZ_BKTsuperfluid} by using the Poisson resummation.

\section{Finite-\texorpdfstring{$D$}{D} effect in the 3D \texorpdfstring{$O(2)$}{O(2)} model in the thermodynamic limit}
\label{appendix:3DXY}

\begin{figure}
    \centering
    \includegraphics[width=12cm]{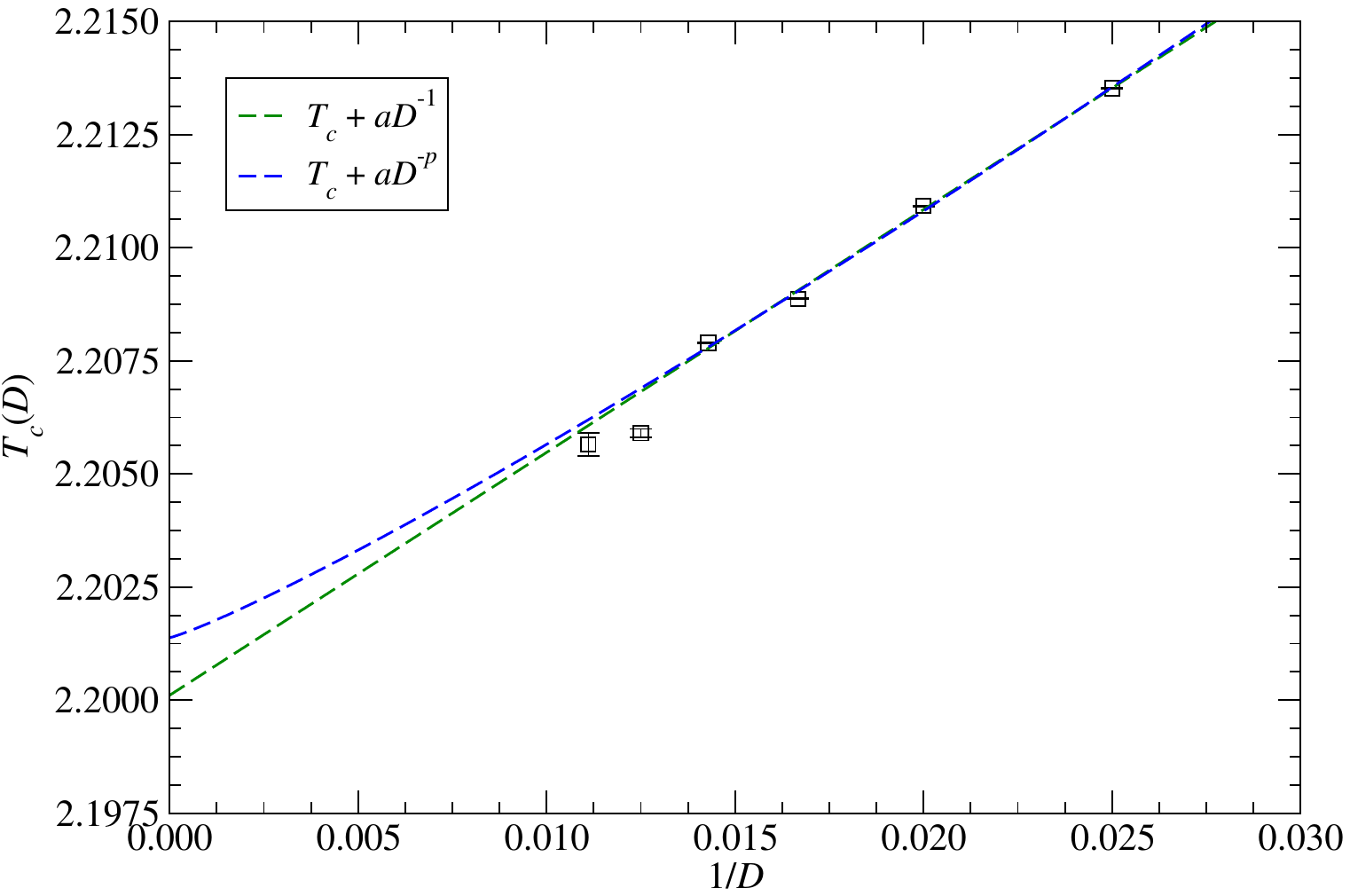}
    \caption{
        Critical temperatures determined from $Z_{\alpha=\pi}/Z_{0}$ in the thermodynamic limit at several finite bond dimensions for the 3D $O(2)$ model. The green dashed line represents a fit using $T_{c}(D)=T_{c}+a/D$, while the blue one denotes a fit using $T_{c}(D)=T_{c}+a/D^{p}$.
    }
    \label{fig:tc_D_3DXY}
\end{figure}

In this Appendix, we briefly discuss the finite bond-dimension effect of the 3D $O(2)$ model when we take the thermodynamic limit. Here, we employ the ATRG algorithm.  

Since $Z_{\alpha=\pi}/Z_{0}$ becomes a step function in the thermodynamic limit, as shown in Eq.~\eqref{eq:behavior_of_Zg}, the critical temperature can be identified as the point at which $Z_{\alpha=\pi}/Z_{0}$ jumps from 0 to 1.
Fig.~\ref{fig:tc_D_3DXY} shows the resulting estimates of the critical temperature obtained for several bond dimensions $D\le90$.
These transition temperatures $T_c(D)$ are all above $2.2050$, which are significantly larger than the critical temperature $T_c=2.2017(2)$ obtained in the main text. 
This clearly shows that the bond-dimension cutoff has a huge effect in the large-volume limit for the 3D $O(2)$ model. 

In the main text, we have circumvented this problem by restricting our considerations for relatively small volumes, $L=2,4,8,16$, which has another advantage that enables us to extract the critical exponent $\nu$ via finite size scaling. However, let us check if we can obtain consistent results for $T_c$ directly in the thermodynamic limit by taking the large-$D$ extrapolation. 
%To estimate the critical point, we extrapolate $T_{c}(D)$ to the $D\to\infty$ limit.
Assuming the $D$-dependence of the form $T_{c}(D)=T_{c}+a/D$, we obtain $T_{c}=2.20011(2)$ and $a=0.537(1)$.
We also fit $T_{c}(D)$ using a more general ansatz $T_{c}(D)=T_{c}+a/D^{p}$, in order to estimate the uncertainty associated with the fitting ansatz.
This fit yields $T_{c}=2.2014(2)$, $a=0.82(7)$, and $p=1.14(3)$.
Since the functional form of the finite-$D$ correction is not known a priori, we take the result of the latter fit as the central value and estimate the systematic uncertainty from the difference between the two extrapolations. 
This procedure yields $T_c=2.2014(2)(13)$,
which is consistent with our estimate $T_{c}=2.2017(2)$ obtained from the finite-size scaling analysis in Sec.~\ref{subsec:3DXY}.

\bibliographystyle{utphys.bst}
\bibliography{bib/ref,bib/algorithm_TNRG,bib/formalism_TNRG,bib/review_TNRG,bib/conti_dof_TNRG,bib/disc_dof_TNRG,bib/fermi_dof_TNRG,bib/gauge_TNRG,bib/paper,bib/ref_twist,bib/data}
\end{document}